\documentclass[print]{aa}
\usepackage[varg]{txfonts}
\usepackage{lscape}

\begin{document}
\title{A new survey of cool supergiants in the Magellanic Clouds}
\author{Carlos Gonz\'alez-Fern\'andez\inst{\ref{inst1}}
\and Ricardo Dorda\inst{\ref{inst2}}
\and Ignacio Negueruela\inst{\ref{inst2}}
\and Amparo Marco\inst{\ref{inst2}}}
\institute{Institute of Astronomy, University of Cambridge, Madingley Road, Cambridge CB3 0HA\label{inst1}
\and Departamento de F\'{\i}sica, Ingenier\'{\i}a de Sistemas y Teor\'{\i}a de la Se\~nal, Universidad de Alicante, Apdo. 99, 03080 Alicante, Spain\label{inst2}}

\abstract{}{In this study, we conduct a pilot program aimed at the red supergiant population of the Magellanic Clouds. We intend to extend the current known sample to the unexplored low end of the brightness distribution of these stars, building a more representative dataset with which to extrapolate their behaviour to other Galactic and extra-galactic environments.}{We select candidates using only near infrared photometry, and with medium resolution multi-object spectroscopy, we perform spectral classification and derive their line-of-sight velocities, confirming the nature of the candidates and their membership to the clouds.}{Around two hundred new RSGs have been detected, hinting at a yet to be observed large population. Using near and mid infrared photometry we study the brightness distribution of these stars, the onset of mass-loss and the effect of dust in their atmospheres. Based on this sample, new a priori classification criteria are investigated, combining mid and near infrared photometry to improve the observational efficiency of similar programs as this.}{}
\maketitle

\section{Introduction}

When stars with masses between $\sim8$ and~$\sim25\,M_{\sun}$ deplete the hydrogen in their cores, they quickly evolve away from the main sequence to the cool side of the Hertzsprung-Russell diagram, becoming red supergiants (RSGs). Evolutionary models indicate that this change happens at almost constant bolometric luminosity. Therefore the decrease in temperature has to be compensated by a rapid and very significant expansion of the atmosphere, which reaches a radius in the range of 400 to 1500~$R_{\sun}$.

The age range for stars in the RSG phase goes from $8$~Myr to~$20$~Myr \citep{eks2013}. All RSGs are thus young stars, associated with regions of recent star formation. However, the nature of the initial mass function constrains the number of stars able to evolve into RSGs: \cite{cla2009b} estimated that a cluster must have a minimum initial mass approaching $10^{4}\,M_{\sun}$ to guarantee the presence of 2\,--\,4 RSGs at a given time.

The RSG phase is critical to our understanding of high-mass star evolution. This phase lasts $\lesssim10$\% of the lifetime of a star, but the physical conditions, specifically mass loss, as a RSG will determine its final evolution. In consequence, evolutionary models for high-mass stars find a strict test-bed in the RSG phase. However, the physics of RSGs defies the limits of 1D computations, and much more complex models are needed to explain observations \citep{eks2013}.

The interest of RSGs goes beyond their role as evolutionary model constraints. Their high luminosity, from $\ga10^{4.5}$ to $10^{5.8}\:L_{\sun}$ \citep{hum1979}, combined with the fact that their emission peaks in the near infrared (NIR) allows their observation at these wavelengths out to very large distances, even if they are affected by high extinction. Thanks to this, in the past few years several massive and highly reddened clusters have been discovered in the inner Galaxy \citep{cla2009a,dav2007,dav2008,dav2012,neg2010}. As their only observable components are the RSGs, these clusters are known as red supergiant clusters (RSGCs).

Recently, \cite{neg2011,neg2012} searched for new RSGs in the region where the end of the long galactic bar is touching the base of the Scutum-Crux arm \citep{dav2007}. This region contains at least three large RSGCs, but as these works show, there are many other RSGs around these clusters, all with similar radial velocities, suggesting that they all belong to the same dynamical structure. For this search, many candidates were selected based on photometric criteria, but other late-type luminous stars, such as red giants or asymptotic giant branch (AGB) stars, have similar photometric characteristics. Therefore the only way to disentangle these populations is by looking at their spectra.

The aim of the present study is to extend our search for new RSGs to other galaxies. Contrasting extragalactic samples with those of the Milky Way will provide information about the processes of stellar formation and evolution. Also, \cite{dav2010} showed that RSGs can be used as abundance probes, opening up a new method to study the metallicity in other galaxies. Finally, the physical characteristics of RSGs vary for different metallicities \citep{hum1979a,eli1985,lev2006,lev2012}, but the number of known RSGs beyond the Galaxy and the Magellanic Clouds (MCs) is very low \citep{lev2013}, and there is not a complete sample of RSGs in any Galaxy, not even the relatively nearby MCs.

In this extragalactic search, our first step are the MCs, where a large RSG population has already been studied  \citep{hum1979a,hum1979b,pre1983,eli1985,lev2006}. This has the big advantage that the distances to both clouds are well established, removing part of the degeneracies that plague the studies of RSGs in the Milky Way. However, until now, only the brightest RSGs of the MCs have been studied, leaving almost unexplored the dimmer end of the RSG population and the frontier between RSGs and AGBs.

In the last fifty years many works have studied photometrically the red population of the MCs, the high-mass population, and the RSGs themselves. The most recent and exhaustive among these works are \citet{mas2002} and \citet{bon2009,bon2010}. \cite{mas2002} surveyed the MCs in the visible, looking for high-mass stars. He found $\sim$280 RSGs in the Large Magellanic Cloud (LMC) and $\sim$160 in the Small Magellanic Cloud (SMC), many of them previously unknown. However, as RSGs have their emission peak in the near infrared and their high mass-loss tends to redden them, by using visible data only the the less reddened and/or optically-brightest RSGs were observed.

Bonanos et al.\ explored the mid infrared (MIR) properties of high-mass stars in the LMC \citep{bon2009} and SMC \citep{bon2010}. Using spectral classifications taken from the literature, they derived new MIR criteria to identify RSGs. \cite{bri2014} used these criteria to select a few dozen candidates in other galaxies, finding six new RSGs.

There have also been some spectroscopic surveys aimed at these populations. The first works \citep{hum1979a,eli1985} studied a small number of the brightest RSGs, confirming their nature. More recent works have taken advantage of the availability of large scale photometric surveys to select large numbers of candidates that can be observed efficiently using multi-object spectroscopy. \cite{mas2003b} obtained spectra for a statistically significant sample taken from their previous photometric survey, confirming the RSG nature of most of their candidates. \cite{neu2012} did a selection of RSGs and yellow supergiant (YSG) candidates using 2MASS. Of their 1949 RSG candidates, they observed 522, labelling 505 of them as "probable supergiants", as even when their apparent magnitudes and radial velocities were compatible with membership to the LMC, the authors did not perform any spectral classification.

In this work, we present a larger sample of candidate RSGs in the MCs, all of which receive a spectral classification based on intermediate-resolution spectra. We analyse the success rate of different photometric selection criteria and discuss the possibility of generating clean samples of RSGs. In addition, we take advantage of photometric catalogues to study the spectral energy distributions of all candidates and search for further selection criteria. Finally, we discuss statistical properties of the RSG population in the MCs.

\section{Target selection and observations}

\subsection{Overall strategy}
Observations were carried out with AAOmega at the Anglo-Australian Telescope as backup/filler for a longer programme aimed at the inner disc of the Milky Way during 9 nights between 2010 and 2013. The observational strategy behind this complementary program was built upon three different samples: a set of photometrically selected candidates, a group of previously known RSGs \citep[from][]{eli1985,mas2003b} and a third group of known YSGs from \cite{neu2010} used as low priority targets in areas of low target density (only in the SMC, as in the LMC we constrained ourselves to an area of high target density).

The samples of already known supergiants (SGs) were added partly as a control sample to compare with our new candidates and partly to extend classification criteria and schemes from the blue and optical into the \ion{Ca}{ii} triplet range at low metallicity. This paper deals mainly with the new candidates, while the full potential of the other samples will be exploited in future publications.

\subsection{Selection criteria}
\label{selcrit}
The recent boom in the detection of RSG rich clusters is mainly due to the availability of all-sky, near infrared data, because in this wavelength regime the peculiar extended atmospheres of these stars stand out. In fact, just by using 2MASS photometry \citep{skr2006}, it is possible to define a pseudo-colour $Q=(J-H)-1.8\times(H-K_\mathrm{S})$ able to separate between blue and red stars. This parameter has been proved to be a excellent tool to pick out RSGs, as they often show values of $Q$ similar to those of yellow stars ($Q\sim0.2-0.3$). This property, combined with their unusual brightness in the $K_\mathrm{S}$ band, allows the definition of purely photometric filters to select this population while minimizing interlopers. These criteria, combined with spectroscopic follow-up to confirm the nature of the candidates, have been used successfully and extensively in our own galaxy \citep{neg2011,neg2012,cgf2012} and in this work we extend their use to the Magellanic clouds.

The fact that these dwarf galaxies are not part of the Milky Way has the advantage that foreground objects are more easily filtered out, particularly if they have large measured proper motions. Background objects, in contrast with the disc of our galaxy, are scarce and fall outside the parameter space occupied by RSGs. This comes at the price of a larger distance modulus, but as RSGs are intrinsically very bright, this is not an important issue. Also, the reddening towards the clouds is relatively small, with typical values around $E(B-V)\sim0.1$ \citep{sos2002,kel2006} and so the pseudo-colour $Q$, that relies on the assumption of a given extinction ratio between bands, will not be affected by variable or non-standard extinction laws, at least outside the most reddened sites of recent stellar formation.

\begin{figure}
\centering
\resizebox{\columnwidth}{!}{\includegraphics[width=8.9cm,clip,angle=180]{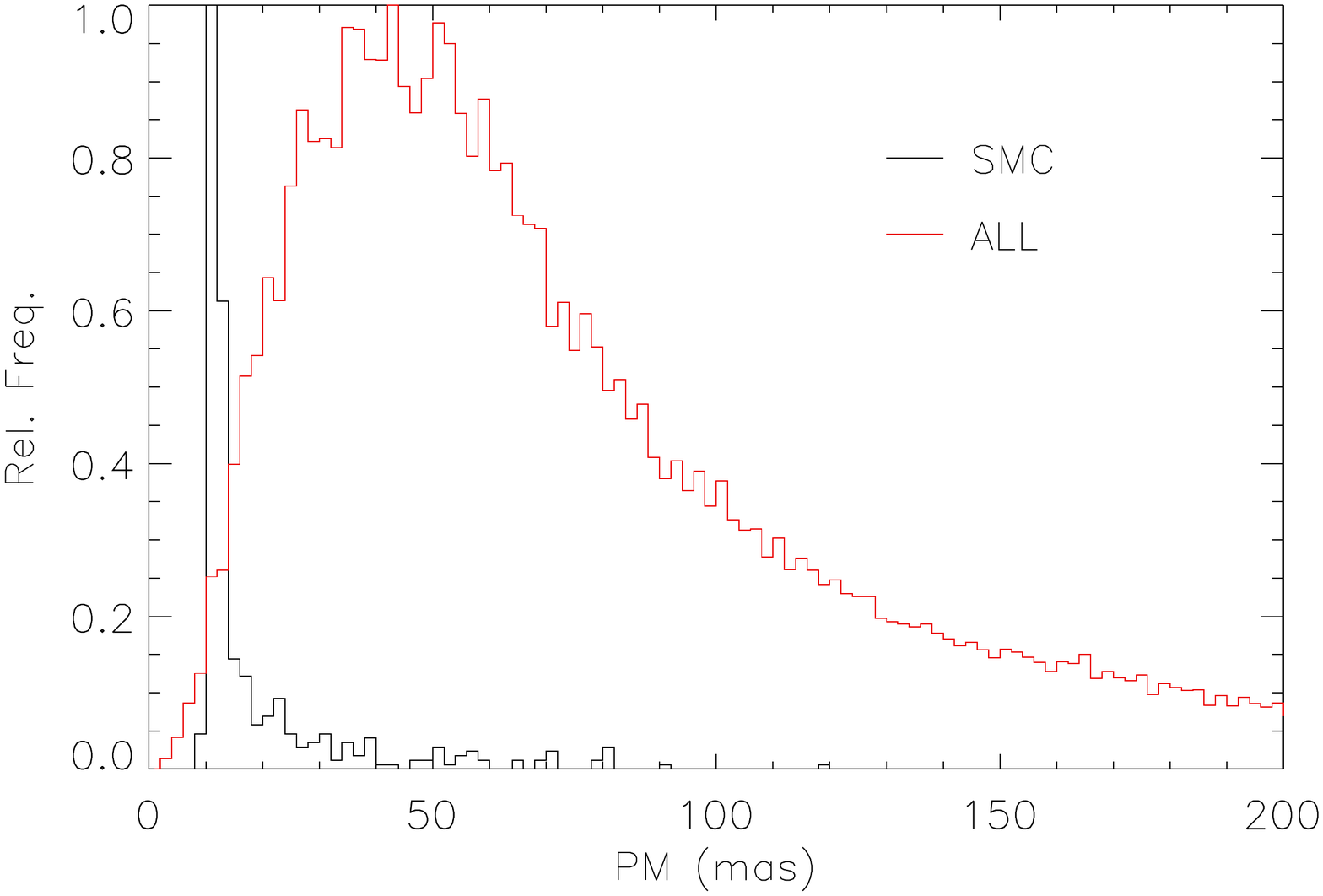}} \caption{Proper motion (taken from USNO-B1) relative distribution for putative RSGs of the SMC (black line) and for all the stars in the field (red line).}\label{pmdist}
\end{figure}

With all these considerations in mind, we defined a set of selection criteria for RSG candidates as follows:
\begin{itemize}
\item[-]Candidates should have $0.1\leq Q\leq0.4$.
\item[-]They have proper motions compatible with the MCs (see Fig. \ref{pmdist}).
\item[-]The brightness divide between RSGs and AGBs is not well established, but RSGs show normally absolute magnitudes brighter than $-8$ in the $K_\mathrm{S}$ band. Taking into account the distance modulus to the clouds, these stars should appear brighter than $K_\mathrm{S}=11$.
\item[-]Lastly, to optimize the observing time, we impose a cut at $m_I=13$ so that spectra with high enough signal-to-noise can be obtained in less than 30 minutes.
\end{itemize}

\subsection{Observations}
While traditional spectral classification criteria for stars are normally defined over the blue end of the optical range, many works have extended them for RSGs to the wavelength range around the infrared \ion{Ca}{ii} triplet, as it contains several atomic and molecular lines of physical interest. With the fibre-fed dual-beam AAOmega spectrograph it is possible to cover both regions of the spectrum for several hundred objects in a single exposure, making it an ideal instrument for this kind of studies. As it sits on the 3.9~m Anglo-Australian Telescope (AAT) at the Australian Astronomical Observatory, it has good access to the low latitude fields of the Clouds while offering a collector area large enough to observe their RSGs in a very efficient manner.

The instrument is operated using the Two Degree Field ("2dF") multi-object system as front end, allowing the simultaneous acquisition of spectra through 392 fibres. These fibres have a projected diameter of $2\farcs1$ on the sky and are fed into the two arms via a dichroic beam splitter with crossover at 5\,700\AA{}. Each arm of the AAOmega system is equipped with a 2k$\times$4k E2V CCD detector (the red arm CCD is a low-fringing type) and an AAO2 CCD controller. While the red arm was always equipped with the 1700D grating, the blue arm changed between the 580V and 1500V gratings. A summary of the configurations is offered in Table~\ref{obsconf}. However, since the projection of the spectrum from each fibre on the CCD depends on its position on the plate, it is not possible to give a precise common range for each configuration. This effect displaces the spectral range limits by $\sim$20\,\AA{} in the red range, $\sim$40\,\AA{} for the 580V grating in the blue range, and $\sim$20\,\AA{} for the 1500V grating in the blue range.

\begin{table*}
\caption{Summary of the observations}
\label{obsconf}
\centering
\begin{tabular}{c c | c c c | c c c}
\hline\hline
\noalign{\smallskip}
&&\multicolumn{3}{c|}{Blue arm}&\multicolumn{3}{c}{Red arm}\\
Year&Nights&Grating&$\lambda_{{\rm cen}}$ (\AA{})&Range (\AA{})&Grating&$\lambda_{{\rm cen}}$ (\AA{})&Range (\AA{})\\
\noalign{\smallskip}
\hline
\noalign{\smallskip}
2010&3&580V&4500&$\sim$2100&1700D&8600&$\sim$500\\
2011&2&1500V&4400&$\sim$800&1700D&8700&$\sim$500\\
2012&1&1500V&5200&$\sim$800&1700D&8700&$\sim$500\\
2012&2&580V&4800&$\sim$2100&1700D&8700&$\sim$500\\
2013&1&580V&4800&$\sim$2100&1700D&8700&$\sim$500\\
\noalign{\smallskip}
\hline
\end{tabular}
\end{table*}

The nominal resolving powers ($\lambda/\delta\lambda$) at blaze wavelength for the 580V and 1500V gratings are 1\,300 and 3\,700, while the 1700D grating reaches $R\sim10\,000$ around the Ca triplet, allowing the measurement of line-of-sigh velocities with enough precision for our purposes.

\begin{figure}
\centering
\resizebox{\columnwidth}{!}{\includegraphics[width=8.9cm,clip]{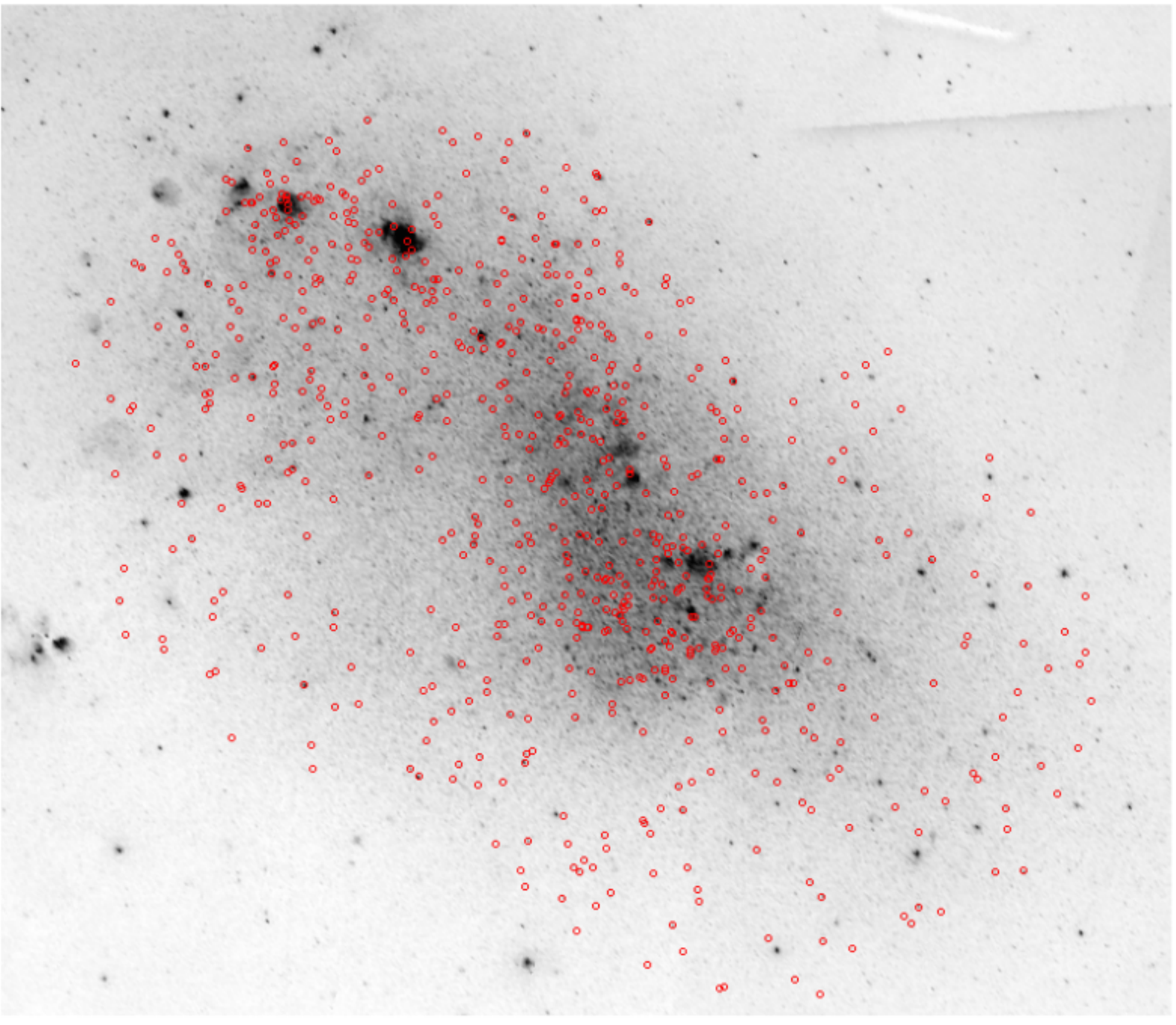}} \caption{Spatial distribution of targets in the SMC, over a DSS-Red image roughly $3^\circ\times3^\circ$ in size.}\label{fovsmc}
\end{figure}

\begin{figure}
\centering
\resizebox{\columnwidth}{!}{\includegraphics[width=8.9cm,clip]{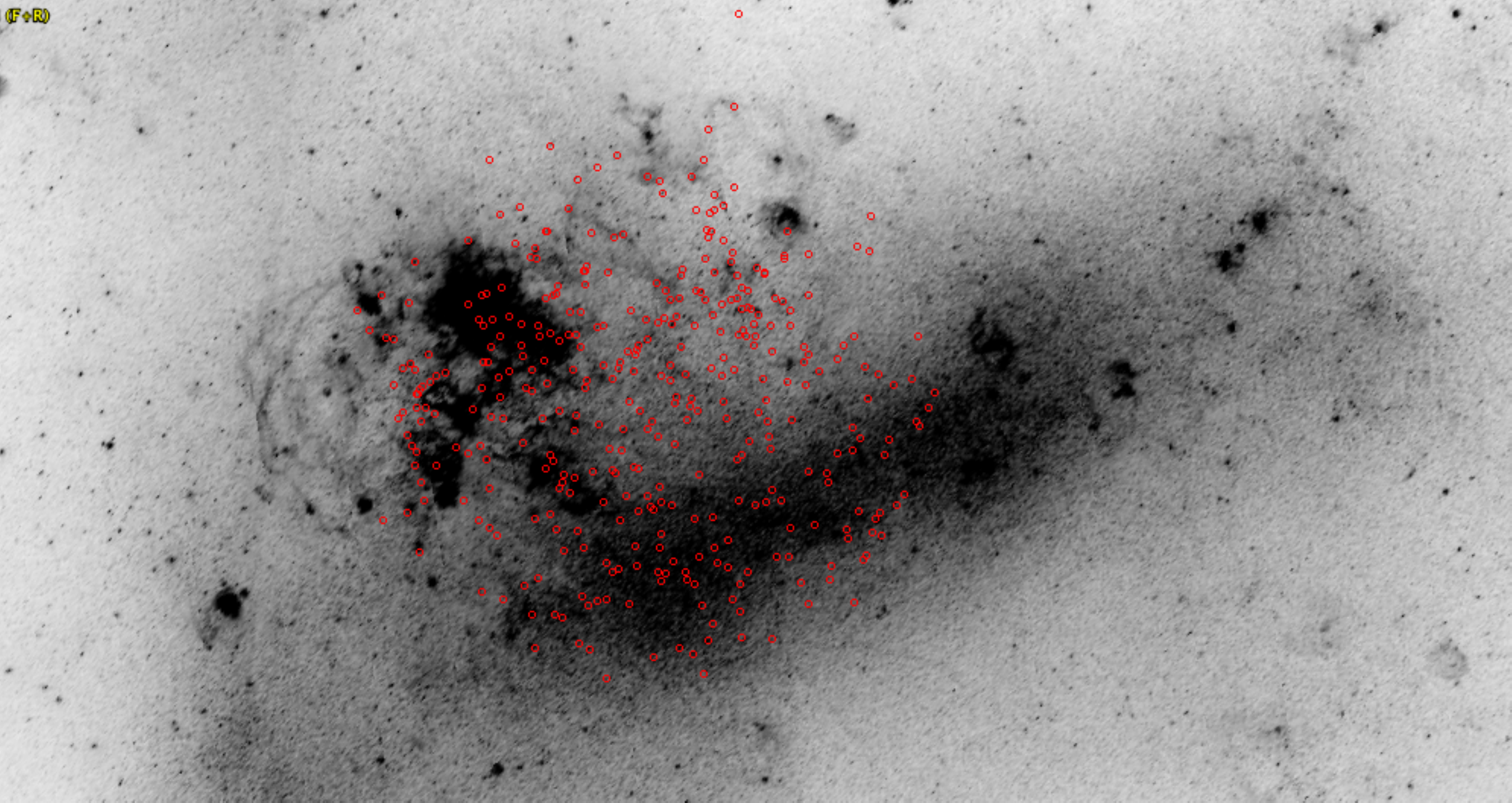}} \caption{Spatial distribution of targets in the LMC, over a DSS-Red image roughly $4^\circ\times2^\circ$ in size. As can be seen, all the targets (from a single AAOmega pointing) are distributed over a region that covers less than 50\% of the galaxy. }\label{fovlmc}
\end{figure}

The main body of the SMC was covered with two pointings (Fig. \ref{fovsmc}) that were observed using 8 different configurations, for a total of 1448 spectra. Only one pointing was devoted to the LMC (Fig. \ref{fovlmc}), visited with two configurations for a total of 464 spectra. As a subset of targets were observed using several configurations and some spectra did not reach usable S/N, our sample amounts to a total of 617 individual objects for the SMC, and 314 for the LMC.

\subsection{Data reduction}
Data reduction was performed using the standard automatic reduction pipeline {\tt 2dfdr} as provided by the AAT at the time. Wavelength calibration was achieved with the observation of arc lamp spectra immediately before each target field. The lamps provide a complex spectrum of He+CuAr+FeAr+ThAr+CuNe. The arc line lists were revised, and only those lines actually detected were given as input for {\tt 2dfdr}. This resulted in very good wavelength solutions, with rms always $<0.1$ pixels.

Sky subtraction was carried out by means of a mean sky spectrum, obtained by averaging the spectra of 30 fibres located at known blank locations. The sky lines in each spectrum were evaluated and used to scale the mean sky spectrum before subtraction.

\subsection{Measuring $v_{{\rm los}}$}
We measured velocities along the line of sight by calculating the correlation function of our observed spectra with a suitable template of known non-cosmological redshift. For late-type stars as the ones here studied, a high resolution spectrum of Arcturus is normally used, but while for the part of the spectrum with $\lambda>1\:\mu$m this is an adequate standard, as the overall shape of the spectrum does not change dramatically, this is not the case for the wavelength range around the Calcium triplet, as in this region the spectrum of successive populations will be dominated by the Paschen series, then the Calcium atomic lines and lastly TiO molecular bands. This results in a rather dramatic change in typology, making it very difficult to find a one-size-fits-all standard to use as comparison.

We have chosen instead to use a whole family of MARCS synthetic spectra taken from the POLLUX database \citep{pal2010}. For each observed spectrum, the most similar model is selected doing a first pass over the whole set of synthetic spectra using very rough increments in velocity ($\Delta v_{{\rm los}}=10~\mathrm{km\,s^{-1}}$) and once the best match is selected, a refined value of $v_{{\rm los}}$ is measured calculating the correlation between observation and model using $0.3~\mathrm{km\,s^{-1}}$ increments. These measured velocities where later transformed into the heliocentric system of reference using the {\em rvcorrect} package from {\sc Iraf}.

\begin{figure}
\centering
\resizebox{\columnwidth}{!}{\includegraphics[clip,angle=180]{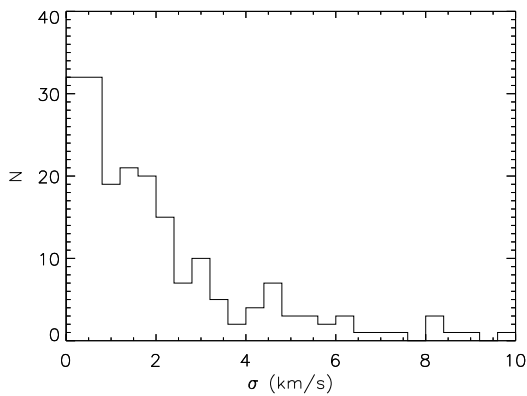}} \caption{Histogram of the standard deviation for the measured $v_{{\rm los}}$ of repeated observations.\label{sigvlos}}
\end{figure}

Using stars with repeated observations, we can obtain an estimate of the total uncertainty in $v_{{\rm los}}$, including measuring errors, wavelength calibration, astrophysical noise, etc. As can be seen in Fig.~\ref{sigvlos}, the typical velocity dispersion is around $1.0~\mathrm{km\,s^{-1}}$, and we can assume a conservative $99\%$ confidence interval for our measurements of $v_{{\rm los}}$ at $4~\mathrm{km\,s^{-1}}$. Another source of dispersion that needs to be taken into account are possible systematic effects between different observing runs. Using all the available stars (main program and SMC/LMC backup) we checked for these, as we have repeated measurements for several objects. Systematic differences in $v_{\mathrm{hel}}$ were all below $1~\mathrm{km\,s^{-1}}$ and have not been corrected, as some of the fields (particularly in the LMC) have very low redundancy and hence is difficult to measure and correct properly for this effect.

In this article we will only use velocities in a relative sense, to discriminate between populations from the MCs and from different Galactic components. Being so, we only worry about the internal consistency of the calibration, without the need of an anchor point to check for systematics. In any case, as can be seen in Fig.~\ref{membervel}, the derived systemic velocity for the SMC is in very good agreement with the values from the literature, hinting at a very low systematic residual, if any. This is not the case for the LMC. Since we are not surveying the totality of the galaxy, we are heavily biased by its internal dynamic structure, and cannot readily compare with an "average" galactic velocity.

\section{Results}

\subsection{Spectral classification}
\label{spcl}
To classify the observed stars we used spectra obtained with the blue arm, as they cover the wavelength range where classical classification criteria are defined \citep{mor1943,fit1951,kee1976,kee1977,kee1980,mor1981,tur1985,kee1987}. These criteria were complemented with our own secondary indicators, whose variation with spectral type (SpT) and luminosity class (LC) we derived through visual comparison between the spectra of known standards. These were taken from the Indo-US spectral library \citep{val2004} and the MILES star catalogue \citep{sbl2006}, and degraded to our spectral resolution (roughly a FWHM of $2.1~$\AA{}) when needed.

\citet{hum1979a} reported that the metallicity differences between the Milky Way (MW), the LMC and the SMC do not change the behaviour of the atomic line ratios and other spectral features used in the spectral classification of RSGs. Therefore, it is possible to use MW standards as comparison, and the same criteria developed for RSGs in one galaxy are applicable to the others, as long as they are based on line (or band) ratios and not line (or band) strengths. With our extended sample, we can confirm that there are no apparent differences in the spectra of RSGs from both clouds, even considering that our sample covers a rather broad spread in spectral types. In consequence, we adopted the same criteria for both MCs, using Galactic standards as comparison.

We have performed our own classification for all the stars observed, even those with early SpTs (most of them part of the YSG control sample). As this work centres around RSGs, we will only discuss the detailed classification of stars with spectral type later than G0. This also avoids the metallicity effects over the classification of earlier spectral types, that is more heavily affected by this parameter \citep[cf.][]{evans03}. We have also found some carbon and S-stars among our targets, but these are easily identified due to their very characteristic spectra. As these are interlopers in our sample of new candidates, we did not perform any further analysis on them.

Spectra observed with the 580V grating cover roughly from $3730~$\AA{} to $5850$~\AA{} (the exact limits depend on fibre position), but the S/N blueward of $\sim4500$~\AA{} is very low for many of our stars (this is not the case for the earlier spectral types). In consequence, most of our classification criteria lie between $4500~$\AA{} and $5850~$\AA{}.

The main LC indicators we used are the ratios between the the lines of the Mg\,{\sc{i}} triplet (5167~\AA{}, 5172 and 5184~\AA{}) \citep{fit1951}. From G0 to $\sim$M3, Mg\,{\sc{i}}~5167 is clearly deeper than the other lines for LC~I. These ratios change slowly with SpT, but this variation is not enough to complicate the identification of mid- and high-luminosity SGs (Iab, Ia). There are other spectral features that can be used to confirm the LC: the ratio of Fe\,{\sc{i}}+Y\,{\sc{ii}} blend at 4376~\AA{} to Fe\,{\sc{i}} at 4383~\AA{} \citep{kee1976}, the ratio between the blended Fe\,{\sc{i}} lines around 5250~\AA{} and the Ca\,{\sc{i}}+Fe\,{\sc{i}}+Ti\,{\sc{i}} blend at 5270~\AA{} \citep{fit1951} and the ratio of Mn\,{\sc{i}}~5433~\AA{} to Mn\,{\sc{i}}~5447~\AA{}.

As the Balmer lines decrease in strength with SpT while at the same time metallic lines become more intense, to identify G or earlier subtypes we used the ratio of the H$\beta$ and H$\gamma$ transitions to other nearby metallic lines. For those stars with enough S/N in this region, we also compared H$\gamma$ to the CH G--band (from 4290 to 4314~\AA{}): F stars have deeper H$\gamma$, at G0 H$\gamma$ and the G--band have similar depths and from G0 to mid G, the G--band becomes dominant.

For later SpT, the shape of the metallic lines changes due to the appearance of TiO bands. The sequence of M types is defined by the depth of these bands \citep{tur1985}, but their effects are noticeable from K0 onwards. Starting at K1, a TiO band rises at 5167~\AA{}, changing the shape of the continuum between Mg lines. We used this change to obtain the spectral type for stars between K1 and M2. However, at $\sim$M3, the band is so deep that the Mg lines are not useful any more. This is also the case for the other indicators mentioned before, whenever there is a band close to them.

The effect of the TiO band at 5447~\AA{} over two lines, Mn\,{\sc{i}}~5447~\AA{} and Fe\,{\sc{i}}~5455~\AA{}, can be used to obtain the subtype of K stars, and the the atomic lines and molecular bands between 5700 to 5800~\AA{}, to obtain the spectral type for mid and late M stars (Mn\,{\sc{i}}~5718~\AA{}, Mn\,{\sc{i}}~5718~\AA{}, V\,{\sc{i}}~5727~\AA{}, VO~5737~\AA{}, TiO~5759~\AA{}, Ti\,{\sc{i}}~5762).

For spectra observed with the 1500V grating, we had to resort to different criteria, but the methodology was the same. We identified the LC using the following ratios from \cite{kee1976}: Fe\,{\sc{i}}+Sr\,{\sc{ii}}~4216~\AA{} to Ca\,{\sc{i}}~4226~\AA{}, Fe\,{\sc{i}}+Y\,{\sc{ii}}~4374.5~\AA{} to Fe\,{\sc{i}}~4383~\AA{} and Fe\,{\sc{i}}~4404~\AA{} to Fe\,{\sc{i}}+V\,{\sc{i}}+Ti\,{\sc{ii}}~4409~\AA{}. In all cases, the ratio is $\sim$1 for LC~I and $\gg$1 for less luminous stars (LC~III\,--\,V).

SpT can be evaluated by comparing H$\delta$ at 4102~\AA{} and H$\gamma$ at 4341~\AA{} with nearby metallic lines. For early-G types the depth of H$\gamma$ is similar to that of the G band, while for F or earlier types, H$\gamma$ is clearly dominant. Even if the Fe\,{\sc{i}}~4347~\AA{}, Fe\,{\sc{i}}+Cr\,{\sc{i}}+Ti\,{\sc{ii}}~4344~\AA{} and Mg\,{\sc{i}}+Cr\,{\sc{i}}+Fe\,{\sc{i}}~4351~\AA{} lines vary with LC, they can be used to determine SpT, because we can constrain this parameter with other indicators.

We also used the ratios between Fe\,{\sc{i}}~4251~\AA{} and Fe\,{\sc{i}}~4254~\AA{}, Fe\,{\sc{i}}~4280~\AA{} and Fe\,{\sc{i}}+Ti\,{\sc{i}}~4282~\AA{}, and the behaviour of the lines Fe\,{\sc{i}}+Co\,{\sc{i}}~4579.5~\AA{}, Fe\,{\sc{i}}~4583~\AA{} and Fe\,{\sc{i}}+Cr\,{\sc{i}}+Ca\,{\sc{i}}~4586~\AA{} to confirm the spectral subtypes in the later~G and K sequences.

Despite the fact that the first TiO bands appear already for K type stars, it is only in redder spectral regions. In the spectral range considered here they are not noticeable until early M subtypes, and no TiO bands are clearly visible before $\sim$M3. To classify the early M stars we used the spectral range from 4580 to 4590~\AA{} and from 4710 to 4720~\AA{}.

In order to facilitate calculations, we parametrized SpT and LC over a linear scale, assigning integers to each type and class. In those cases in which we doubt between two consecutive classifications, we assigned the intermediate half-integer value.


Even though the term RSG is normally applied to SGs with types K or later, we have also included G stars for two reasons: firstly, RSGs are intrinsically variable and some can change their type from early K to late G. Secondly, at lower metallicities, the typical SpT of a RSG becomes earlier. In consequence, as \cite{lev2013} noted, if we exclude G stars we are losing part of our target population (evolved high mass stars), specially for low metallicity galaxies such as the SMC. Therefore we have used all the stars with SpT G0 or later for subsequent calculations.


There is some overlap between different SMC observations. As we performed the classification for each of the spectra of these redundant targets independently, we can use use them to test the internal coherence of our classification scheme. The final SpT and LC for these stars were obtained by averaging and using the S/N of each spectrum as weight, rounding the final figure to the closest entire or semi-entire number.

The mean differences between the spectral classifications of these repeated targets are given in Table~\ref{differences}. As can be seen, the differences in both LC and SpT are of the order of the classification step, as long as we take into account that we assigned semi-entire SpT only in those cases where the classification between two consecutive subtypes was not clear.

Attending to the obtained differences in our classification, we have assumed an uncertainty of $\pm1$ in SpT and $\pm0.5$ in LC for all our stars, even if there are no repeated observations for the LMC, as the observing conditions and the classification scheme were the same for all fields.

\begin{table}
\caption{Repeatability of our spectral and luminosity classification using stars with multiple observations.}
\label{differences}
\centering
\begin{tabular}{c c c | c}
\hline\hline
\noalign{\smallskip}
&$\overline{\Delta (LC)}$&$\overline{\Delta (SpT)}$&Number of stars\\
\noalign{\smallskip}
\hline
\noalign{\smallskip}
2010&0.4&1.2&99\\
2011&0.4&0.8&101\\
2012&0.4&0.9&129\\
\noalign{\smallskip}
\hline
\noalign{\smallskip}
Weighted average&0.4&1.0&329\\
\noalign{\smallskip}
\hline
\end{tabular}
\end{table}

Of all the stars with more than one observation, there are a few that present large discrepancies between epochs. Even if the numbers are compatible with normally distributed errors, we revised all these spectra to check the source of these differences. In many cases it is due to one of the spectra having low S/N. In these cases, as our final classification was done using the S/N as weight, the final result will be dictated by the high S/N classification. Other stars have good S/N in all their spectra, and  differences arise due to the lack of enough standards for some spectral subtypes and luminosity classes. This is the case of many G stars.

A high fraction of RSGs are known to be long period variables \citep{woo1983}. These variations reflect not only on their brightness, but also on their spectra and radial velocities. Therefore, among many of our repeated observations, stars will appear with  different SpT, LC and radial velocities from epoch to epoch. These classifications have also been averaged. As much of what is discussed in later sections deals with single epoch photometry, there is no use on retaining several classifications for the same object, and by averaging we ensure that the final values for SpT and LC will not be at any of the extremes, increasing the chances of better agreement with the asynchronous photometric measurements.

However, for all those stars showing variability, we retain the different classifications, while for non-variable objects a single value is listed. We have considered to be variable all those stars which have a difference in SpT or LC, non attributable to other factors, larger than twice the uncertainty interval: 2 subtypes or 1 luminosity subclass.

Finally, we want to stress that our spectral classifications are merely morphological. When we classify an object as a supergiant, we are simply stating that several significant spectral features in its spectrum look more like those of supergiant standards than those of giant standards. We are not making any assumption about the physics of the stellar interior. Even though there is a generally excellent correlation between spectral type and physical characteristics, this does not always have to be the case. For example, recently \citet{mor2013} have presented evidence that $\alpha$~Her, an M5\,Ib-II MK standard and anchor point of the classification system (because it is the high-luminosity standard with a later spectral type), is an AGB star of only $\sim3\,M_{\sun}$.

\subsection{Membership to the clouds}
\label{clmem}
\begin{figure}
\centering
\resizebox{\columnwidth}{!}{\includegraphics[width=12cm]{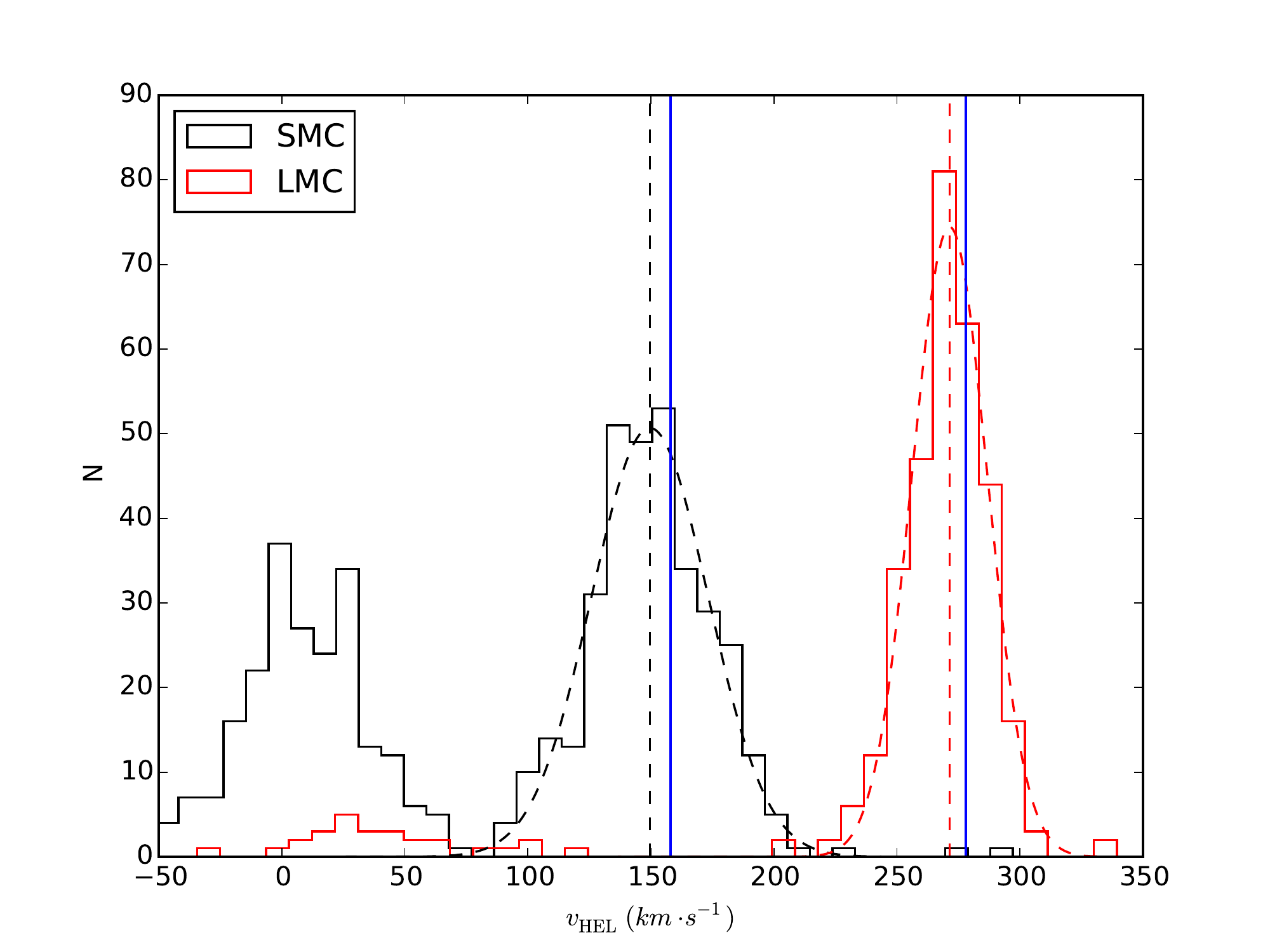}} \caption{Observed heliocentric velocities for all the sources in the LMC (red) and SMC (black). Over-plotted with dashed lines are Gaussian fits to the distributions, and their $\mu$ is marked with a vertical dashed lines. For comparison, the blue vertical lines denote the systemic velocities for the clouds \cite[taken from][]{mas2003b}.}\label{membervel}
\end{figure}

\begin{figure}
\centering
\resizebox{\columnwidth}{!}{\includegraphics[width=12cm]{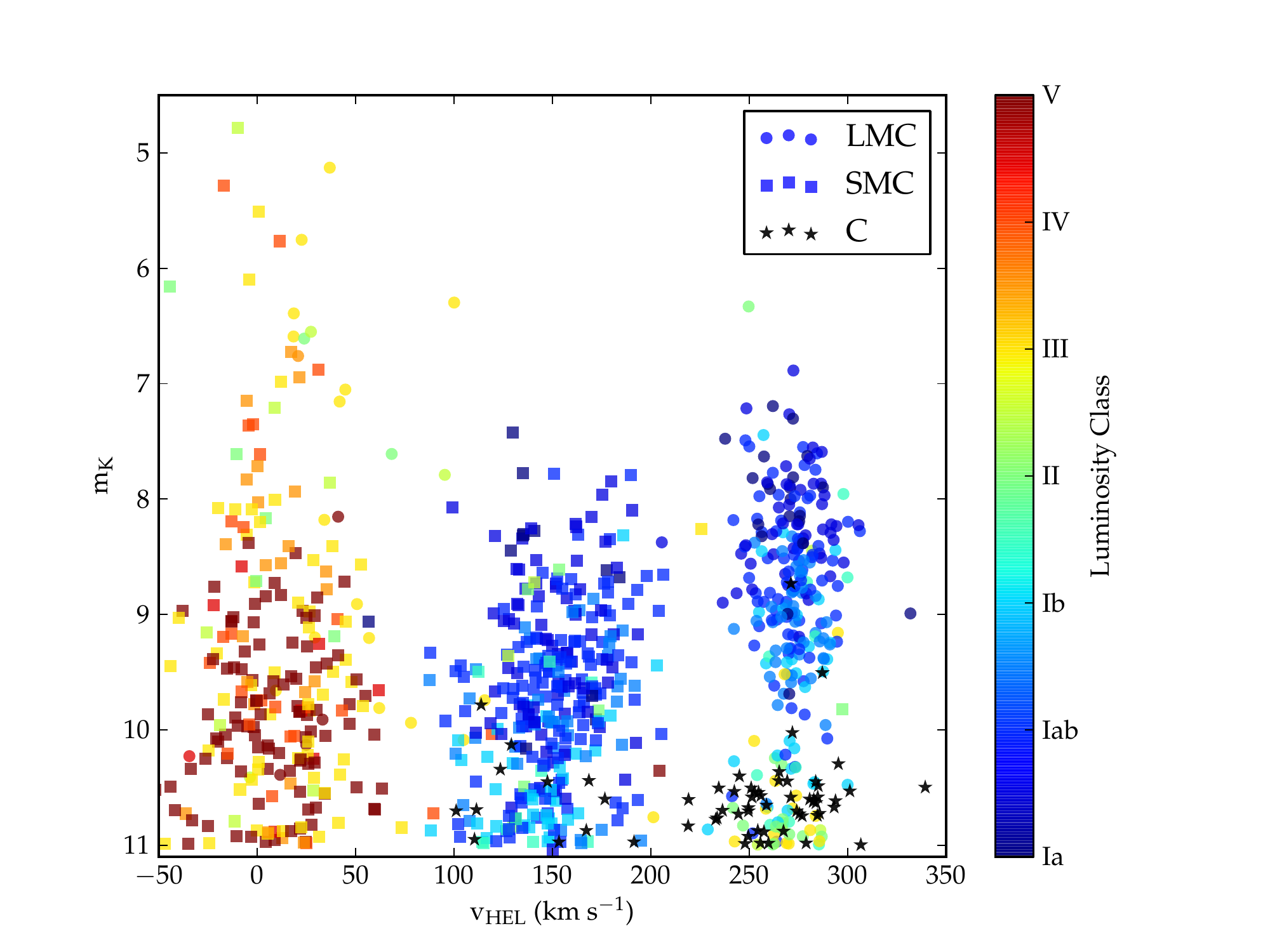}} \caption{Plot of the apparent magnitude versus the $v_{\mathrm{hel}}$ for the observed sources. Squares mark candidates for the LMC, circles do so for the SMC and stars are left for Carbon stars. The colour coding denotes LC, see Sect. \ref{spcl} for an explanation of the chosen parametrization.}\label{vl_lc}
\end{figure}

The velocity distribution of our potential SMC sources can be accurately modelled by a Gaussian distribution with parameters $(\mu=149.6~\mathrm{km~s^{-1}},~\sigma=23.7~\mathrm{km~s^{-1}})$, while for the LMC these become $(\mu=271.4~\mathrm{km~s^{-1}},~\sigma=15.3~\mathrm{km~s^{-1}})$, as can be seen in  Fig. \ref{membervel}. Based on this, an initial clean-up of the sample can be obtained using hard cuts in $v_\mathrm{hel}$, using $\pm3\sigma$ as threshold. Yet the populations from the MW and the MCs cannot be separated based purely on dynamical criteria, as there are halo stars that show velocities compatibles with those of the clouds. This can be seen in Fig. \ref{vl_lc}, where the sources are colour labelled according to their LC. Both for the LMC and the SMC there are stars of classes III to V within the dynamical envelope of the clouds but with apparent magnitudes incompatible with their distance modulus and spectral classification. The only way to weed out these interlopers is through detailed spectral tagging.

On top of these MW populations, the transition from RSG to AGB is smooth, and so both very luminous AGBs and Carbon stars will appear in photometrically selected samples. Although these will indeed be part of the clouds, in order to ensure a pure sample of SGs, it is again mandatory to perform a good spectral characterization of the sources.
Using both $v_\mathrm{los}$ and the spectral classification, we can perform the last cleansing of the sample in order to produce a catalogue of SGs in the MCs. The results from the different stages of this process are shown in Table \ref{filtering}. The final sample contains a total of 160 SGs in the SMC and 123 in the LMC. Of these, 70\% are previously unknown SGs.

\begin{table*}
\caption{Filtering of the original sample of candidates according to several criteria. In parenthesis are the carbon stars with $v_\mathrm{hel}$ not compatible with the clouds and the number of previously undetected SGs.}
\label{filtering}
\centering
\begin{tabular}{c | c c c c | c}
\hline\hline
\noalign{\smallskip}
Cloud&Total&$v_\mathrm{los}$ filter& LC filter&Carbon&Final sample\\
\noalign{\smallskip}
\hline
\noalign{\smallskip}
LMC&237&203&125&48 (3)&123 (70)\\
SMC&400&179&162&10 (0)&160 (128)\\
\noalign{\smallskip}
\hline
\end{tabular}
\end{table*}

\section{Discussion}
\subsection{Selection efficiency and interlopers}
\label{seleff}
In Sect. \ref{clmem} we show that only the combination of $v_\mathrm{los}$ and spectral classification can separate sources from the MW and the MCs. Once this filtering has been done, we can proceed to analyse the efficiency of our selection criteria.

As can be seen in Table \ref{filtering}, of the 585 new candidates, 48\% turned out to have LC Ib-II or brighter, and hence can be classified as SGs. The ratio of success for our selection is of 53\% for the LMC and 45\% for the SMC. This small difference arises mostly due to the fact that we only covered the LMC with one configuration aimed at its main body, while for the SMC we also sampled the outskirts, where the relative density of bright interlopers is higher.

\begin{figure}
\centering
\includegraphics[width=10cm]{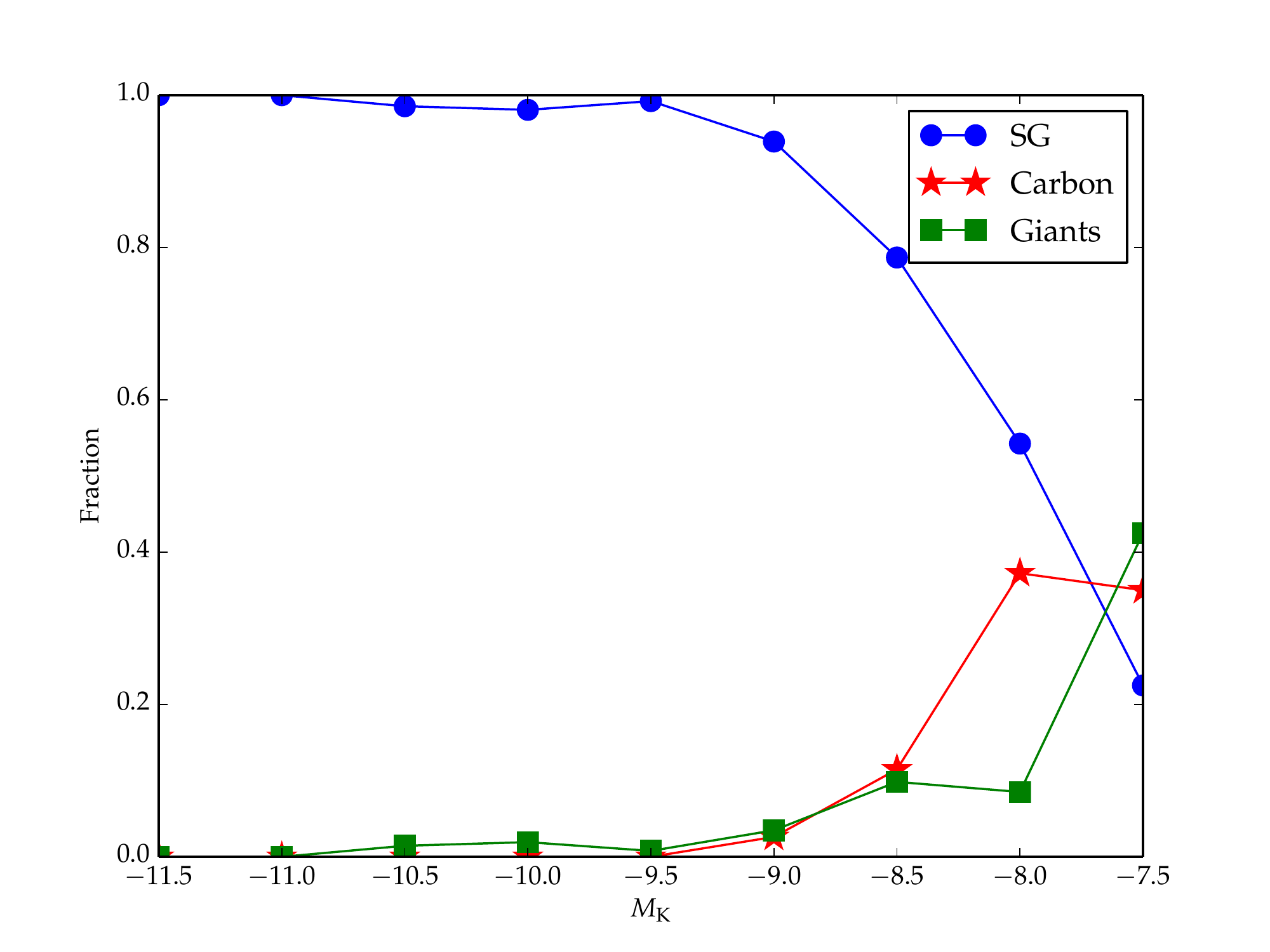} \caption{Fraction of the final sample for both clouds made up by SGs (LC Ib-II or more luminous), giants (LC II or less luminous) and carbon stars. Distance moduli are taken to be 18.48 for the LMC and 18.99 for the SMC, from \citet{walker2012} and \citet{grac2014}}\label{frac_cont}
\end{figure}

Most of these interlopers turn out to be MW disc population, along with some high velocity halo stars. The majority of these could be removed by the application of a cut on proper motions based on a catalogue with more precise measurements than USNO.B-1. Proper motions from {\it Gaia} will allow samples almost clean of MW populations. Among MC interlopers, carbon stars are particularly conspicuous. At low magnitudes, they are the main contaminants in the LMC. Some of these stars may be easily filtered out by looking at their colours, as for the MCs they reach $(J-K_\mathrm{S})>2$, while in our sample no SG is redder than $(J-K_\mathrm{S})\leq1.6$. But since typically, carbon stars will have $(J-K_\mathrm{S})>1.4$ \citep{cole2002}, there is some overlap between both populations, and much more so for fields under heavier extinction, where the SG population will be displaced into the red. This is the case too for mid-infrared colours, where the overlap is even more complete, and from having similar colours, it follows that carbon stars and RSGs will also show similar values of $Q$. Being so, it is expected that these stars will appear in any survey aimed at RSGs with enough depth to reach the low luminosity end of the Ib population, as can be seen in Fig. \ref{frac_cont}: the fraction of recovered SGs drops below $M_\mathrm{K}=-9$, reaching a point at $M_\mathrm{K}\sim-8$ where in fact interlopers dominate the sample.

\subsection{Completeness of the sample}

An item of utmost importance when  considering a sample is that of completeness. In our case, this is delimited by the most restrictive of the criteria outlined in Sect. \ref{selcrit}: the cuts in $Q$. As has been shown, introducing hard thresholds in $Q$ leads to a low proportion of interlopers in the sample, but at the same time there is the chance it might leave out some SGs too. We can check this by comparing our {\it a priori} photometric selection with other similar programs. Of those available, the most complete and deep is \citet{boy2011}, with the added advantage that their classification is based on the NIR and MIR photometry of their sources.

\begin{figure}
\centering
\includegraphics[width=9cm]{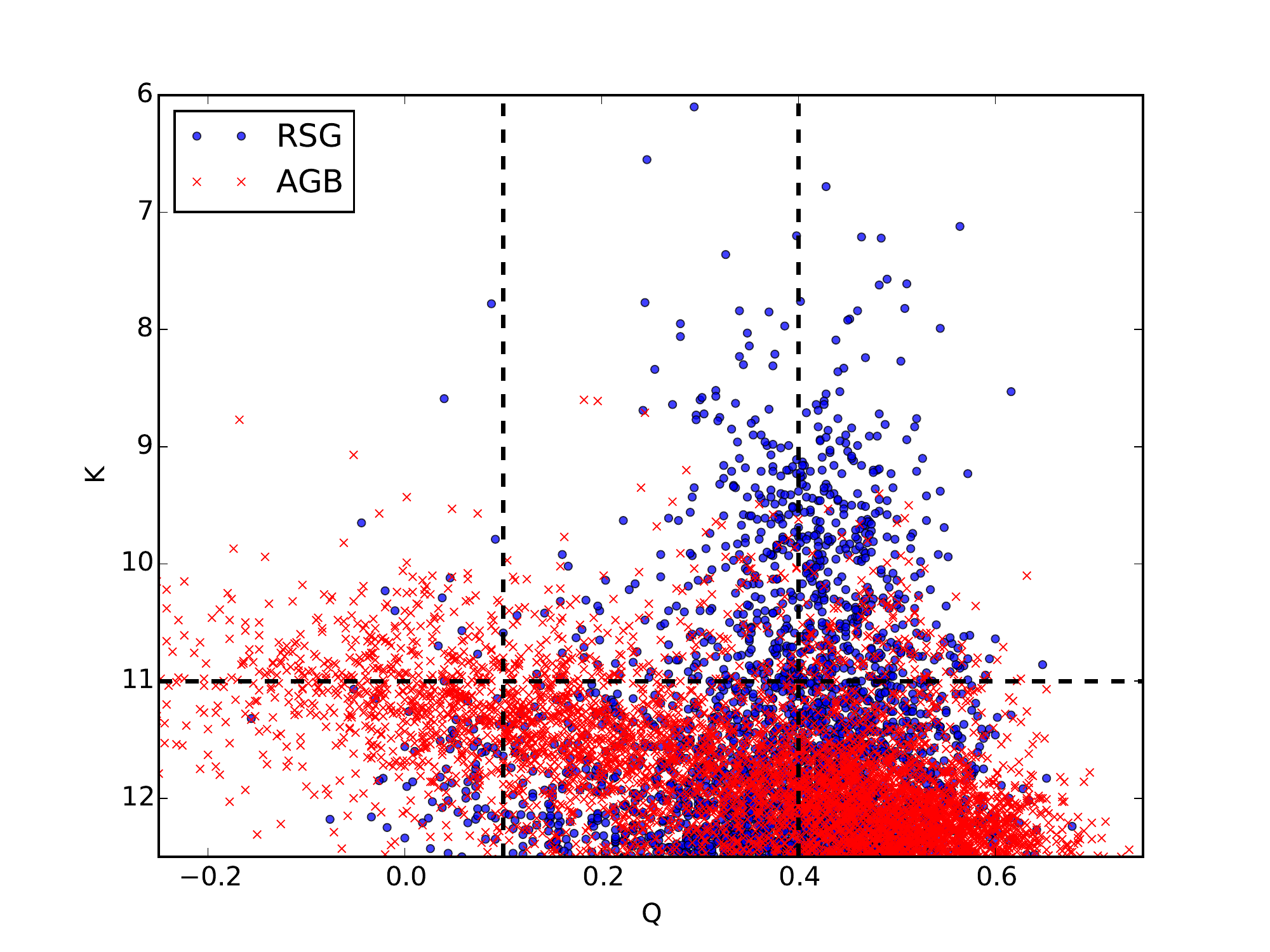}
\caption{$Q$ distribution of the objects from \citet{boy2011}, covering only the SMC. The black dashed lines mark the boundaries of our filtering scheme.\label{boyer}}
\end{figure}

The results of our filtering applied to the objects from \citet{boy2011} can be seen in Fig. \ref{boyer}. At face value, this plot seems to indicate that we are missing  a large fraction of the potential RSGs: of more than 3000 putative candidates flagged by \citet{boy2011} in the SMC, we have only observed around 200. But there are a couple of points that we have to take into account: firstly, stars labelled as RSGs in \citet{boy2011} extend down to $m_{K_{\mathrm{S}}}=12.5$, a magnitude that translates to $M_{K_{\mathrm{S}}}\sim-6.5$ at the distance of the SMC, hardly compatible with what is expected for this population \citep[see, for example,][]{eli1985}; secondly, the spatial extension of the study conducted by \citet{boy2011} is much wider than ours. If we take this into account and we impose that RSGs must have $M_{K_{\mathrm{S}}}<-8$, the equivalent sample from \citet{boy2011} is trimmed down to 479 candidates, of which around 200 have $0.1\leq Q\leq0.4$. A similar result is obtained for other photometric surveys as \cite{mas2002}: of the 288 candidates that have 2MASS photometry, $58\%$ pass the $Q$ filter; and also for \cite{gro2009}, as $40\%$ of their RSGs (stars with $M_\mathrm{bol}\leq-8.0$) clear our cuts. At the other end of completeness, from the 21 RSGs identified by \citet{buch2006} from a pool of objects with colours and fluxes at $8~\mu$m coherent with evolved, massive stars, 17 are picked out by our selection criteria.

As it can be seen in Fig. \ref{boyer}, cutting out sources with $Q>0.4$ does a good job at removing AGBs from the sample (and late type giants, not plotted there), but at the price of also filtering out a large fraction of RSGs. While in the disc of our galaxy, where this kind of filters are mostly used, AGB and giant contamination is a serious concern, and so missing on a significant number of RSGs can be an acceptable price, in the MCs we have the added value of being able to separate a large fraction of these lower luminosity stars just by looking at their apparent magnitudes, so we need to develop finer selection mechanisms.

Beyond its completeness, we can also check if our filtering is biased towards or against given spectral types. Using the control sample, composed of SGs taken from \citet{eli1985}, \citet{mas2003b} and \citet{neu2010}, as these objects were selected disregarding their $Q$ and cover a wide range of spectral types.

\begin{figure}
\centering
\includegraphics[width=9cm]{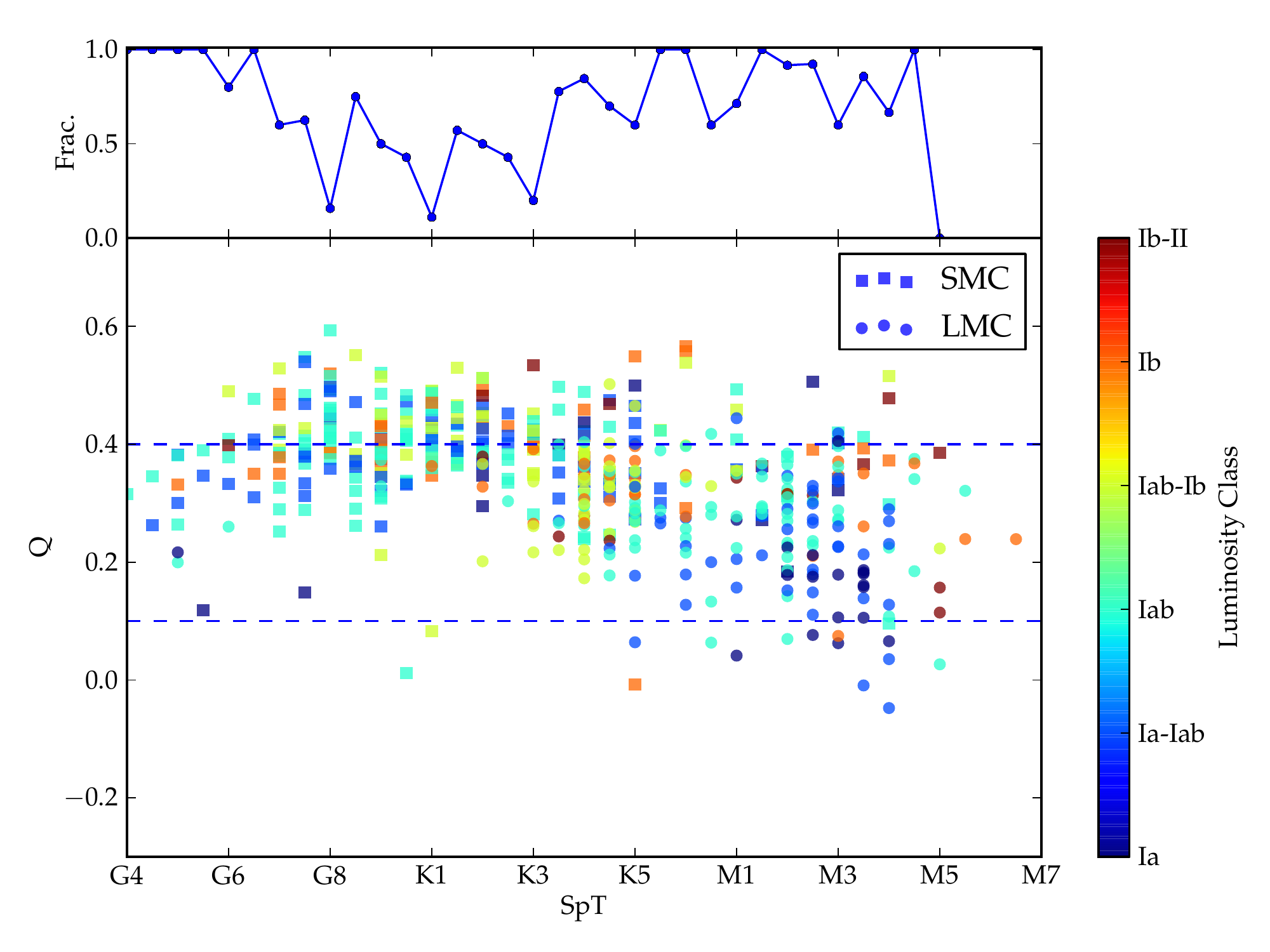} \caption{$Q$ values for the control sample. Top panel: Fraction of these that pass the $Q$ threshold, marked with dashed lines.}\label{qcut}
\end{figure}

\begin{figure}
\centering
\includegraphics[width=9cm]{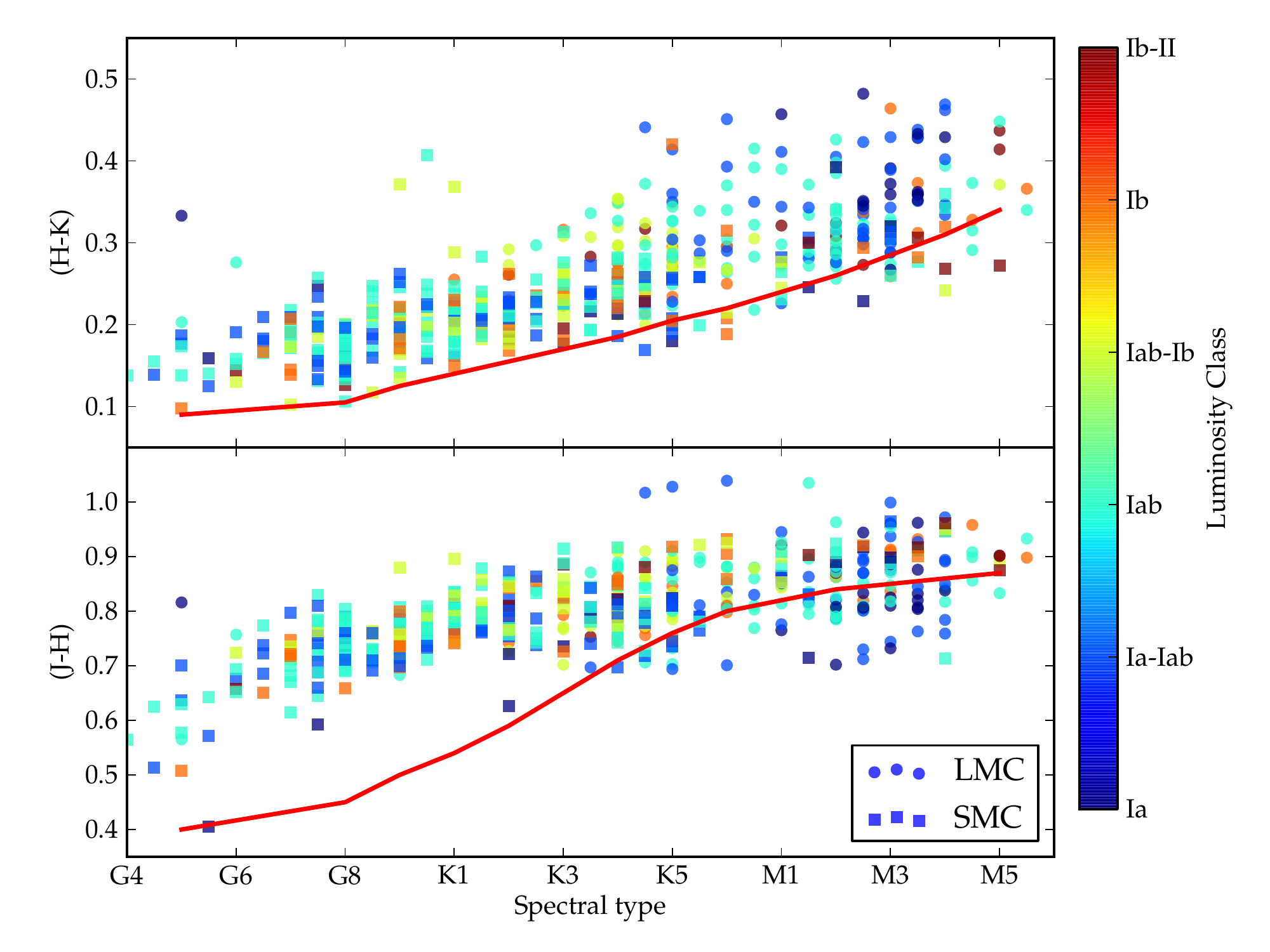} \caption{Evolution of the NIR colours for our control sample as a function of spectral type. Red solid lines mark the intrinsic colours for giants of the same type, taken from \citet{stra2009}.}\label{nircolsp}
\end{figure}

As can be seen in Fig. \ref{qcut},the behaviour of $Q$ with spectral type is relatively complex, related to the variation of the intrinsic NIR colours of SGs. This is shown in Fig. \ref{nircolsp}: the change in $(H-K_\mathrm{S})$ with SpT is similar for giants and SGs (essentially, a temperature sequence), although the latter tend to appear redder. This is not the case for $(J-H)$, where the behaviour is markedly different; this is probably related to the fact that at shorter wavelengths, the spectral energy distribution of SGs is dominated by the combined effects of dust (the interplay between scattering  emission and auto-absorption; \citealt{smith2001}), molecular absorption bands and H$^{-}$ opacity (that drops steadily from $J$ to $H$). These effects dictate the colour of the star, with a weaker dependency on its  temperature.

This implies that our homogeneous filter in $Q$ will have different a priori completeness depending on SpT. This is detailed in the top panel of Fig. \ref{qcut}, where we plot the fraction of SGs from the control sample that clear our filtering: while for mid-type  G  SGs this criterion works very well (even if the relative abundance of these objects is low), its efficiency drops as the spectral type sequence progresses; for late G and early type K SGs it can reach a rather low $\sim30$\% completeness. For later types the fraction of stars inside our boundaries increases more or less linearly with SpT, keeping over $\sim75$\%, although at the very end of the M sequence our sample is too poor to draw any conclusion. This ties back with the results of Fig. \ref{boyer}; as we carried our test on the SMC, where the spectral type distribution works against the efficiency of our $Q$ filtering, the comparison with \citet{boy2011} works as a sort of worst case scenario. Both in the MW and in the LMC the fraction of M type supergiants is much larger, and in particular in the disc of our galaxy the RSG distribution peaks around M2 \citep{eli1985}.

This varying selection efficiency for different spectral types is of paramount importance when devising surveys for galaxies other than our own. It has been shown that the average spectral type of the RSG population depends on metallicity \citep{hum1979a, eli1985, lev2006, lev2012} and so as $Z$ decreases, more RSGs will have earlier types, moving slowly into the region where selection completeness is worse. Although these effects are very difficult to evaluate a priori just based on photometric data, one clear solution is to just open the accepted $Q$ range. As we can see in Fig. \ref{boyer}, this would include in the sample an increasingly large number of interlopers; while for the MCs it would be possible to weed them out, this is not the case in other fields, galactic or extragalactic, and hence the need to develop new strategies arises again.

\begin{figure}
\centering
\includegraphics[width=9cm]{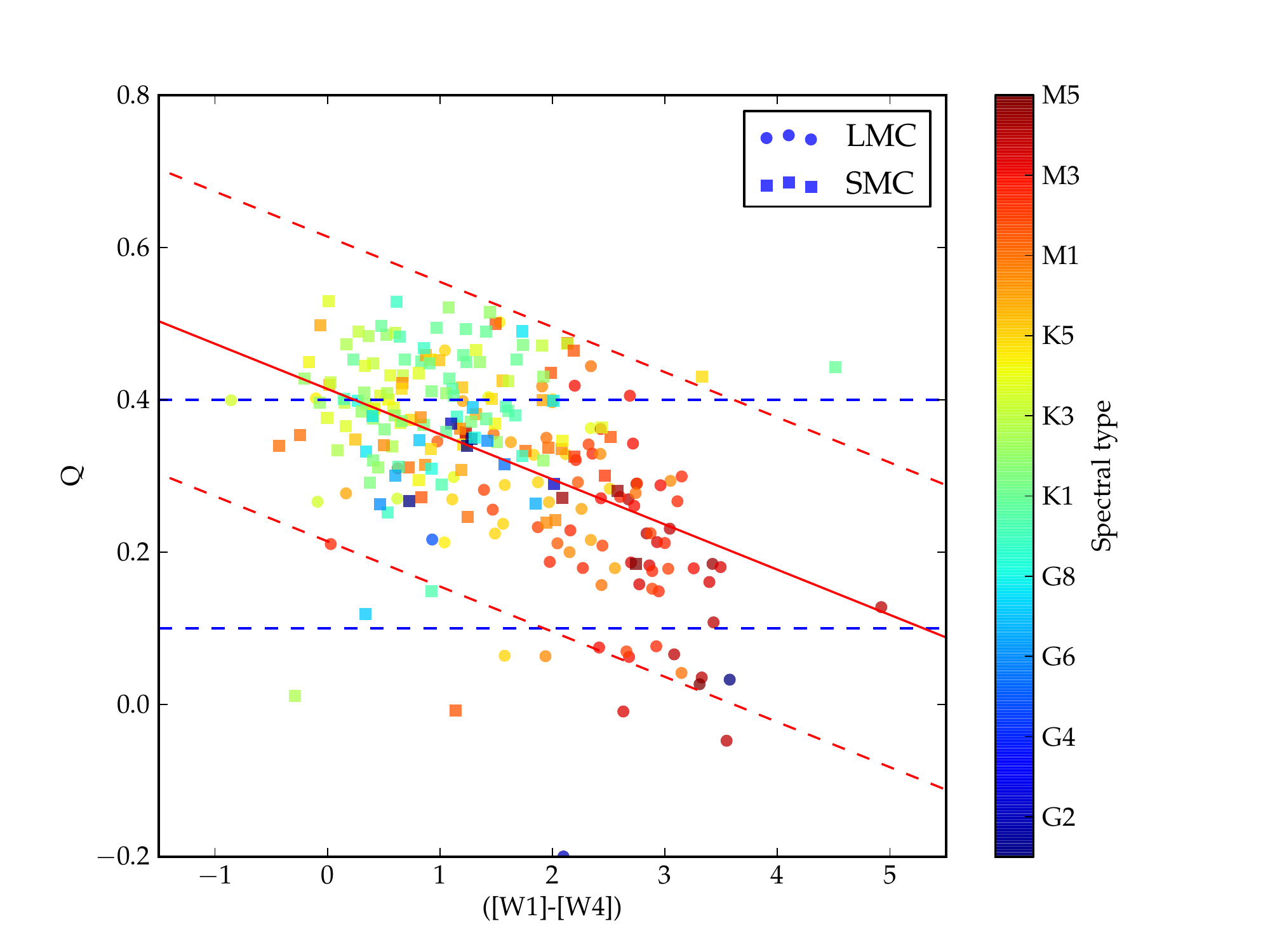} \caption{Pseudo-colour $Q$ as a function of MIR excess for our control sample. The solid red line corresponds to the best linear fit, while the dashed lines denote the $2\sigma$ threshold, that contains 95\% of the SGs.}\label{qmir}
\end{figure}

The influence of dust in the variation of $Q$ is further supported by looking at the MIR colours of these stars, that in WISE \citep{wright2010} show $(W1-W4)$ excesses indicative of dust emission. In fact, $(W1-W4)$ turns out to be a good indicator of spectral types in our sample, and it is possible to refine the selection scheme on $Q$ using this property. As can be seen in Fig. \ref{qmir}, there is a linear relation between $Q$ and $(W1-W4)$, and so we can define an estimation of $Q$ based on MIR colours by $Q_\mathrm{MIR}=0.39-0.06\times(W1-W4)$. Using this, and whenever WISE photometry is available, we can derive a $Q_\mathrm{MIR}$ to compare with $Q$. For our sample, the standard deviation of $Q-Q_\mathrm{MIR}$ is $0.1$, and so it is possible to define a threshold to attain a given completeness. This selection scheme has an important caveat: most late giants will fall within the same range as SGs, and while this is not a problem for other galaxies, where simple cuts in magnitude can weed out these stars, in the MW this strategy is not feasible.

Beyond checking our completeness, we can use these cross-matches with other surveys to evaluate their selection efficiency. This is summed up in Table \ref{cruces}. In this table, we detail those objects in common between the listed surveys and our sample of new RSGs, and whether we confirm their SG nature in the case of photometric surveys or we have discrepant classifications in case of spectroscopic surveys. Two effects mediate the numbers in this table: firstly, we have removed the control sample from this calculation, as these objects are selected a priori knowing their stellar nature. Secondly, the spatial overlap between our survey and those in the table is not complete, and this fact limits the number of common targets.

{\setlength{\tabcolsep}{0.4em}
\begin{table}
\caption{This table summarises the overlap between our sample of new candidates and other works. In the upper panel, we show the objects in common with other purely photometric surveys, and the number of them that we have confirmed spectroscopically as SGs. In the lower panel, we perform the comparison for spectroscopic surveys, indicating the number of SGs and other populations in common. In parenthesis are indicated those cases in which our classification contradicts that of the original work.}
\label{cruces}
\centering
\begin{tabular}{c c c c}
\hline\hline
\noalign{\smallskip}
\multicolumn{4}{c}{Photometric surveys}\\
\hline
\noalign{\smallskip}
Paper&Cloud&Candidates&Confirmed\\
\hline
\noalign{\smallskip}
\cite{wes1981}&LMC&39&29\\
\cite{pre1983}&SMC&36&33\\
\cite{mas2002}&SMC&8&4\\
\cite{mas2002}&LMC&6&5\\
\cite{gro2009}&SMC&7&5\\
\cite{gro2009}&LMC&4&1\\
\cite{boy2011}&SMC&108&99\\
\noalign{\smallskip}
\hline
\hline
\noalign{\smallskip}
\multicolumn{4}{c}{Spectroscopic surveys}\\
\hline
\noalign{\smallskip}
Paper&Cloud&SGs&Other\\
\hline
\noalign{\smallskip}
\cite{eli1985}&SMC&18&0\\
\cite{eli1985}&LMC&7&0\\
OSK1998$^1$&LMC&3&0\\
\cite{mas2003}&SMC&0 (1)&7 (6)\\
\cite{buch2006}&LMC&0 (1)&0\\
\cite{neu2012}&LMC&28&8\\
\noalign{\smallskip}
\hline
\end{tabular}
\tablebib{(1) \cite{oes1998}}
\end{table}
}

\subsection{Photometric properties of the sample}

\begin{figure}
\centering
\includegraphics[width=9cm]{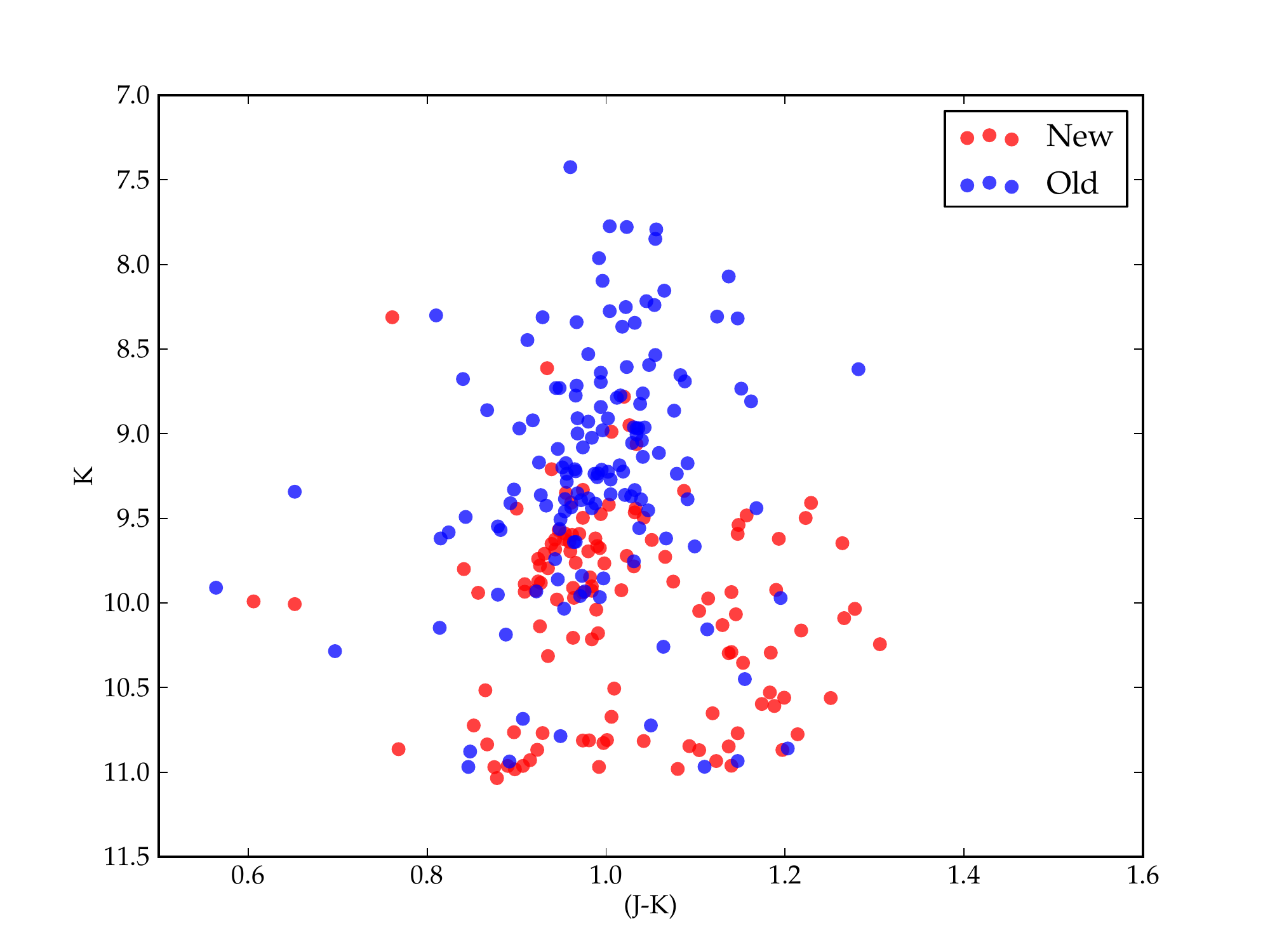} \caption{NIR CMD of all the RSGs in the SMC, where our more complete spatial coverage allows for more meaningful comparison with previous works. Blue dots mark those already present in the literature, while red dots do so for those confirmed in this work. The void around $m_\mathrm{K}\sim10.5$ is an instrumental effect.}\label{oldnew}
\end{figure}

In Fig. \ref{oldnew} we plot a CMD of the confirmed SGs in this work. As can be seen, using the criteria outlined in Sect.~\ref{selcrit} we obtain a set of candidates that, while overlapping with previous works at bright magnitudes, allow us to extend the search for SGs to low brightnesses, in a region of the CMD relatively unexplored. Some of the candidates have already been spectroscopically confirmed as SGs (Table \ref{cruces}), but for homogeneity reasons and due to the variable nature of the spectra of RSGs, they were left as candidates and observed anyway.

Beyond comparing with previous studies, a homogeneous sample this size can be exploited to study the photometric behaviour of RSGs. To do so we combine the photometry coming from 2MASS and WISE (see Table \ref{wvl} for a summary of the wavelength coverage of these surveys). It is worth noting that these are both single epoch surveys, and a fraction of the stars on our sample (AGBs, RSGs, etc.) are expected to be variable. This implies that not only photometry and spectral classification will be asynchronous, but also different bands  have been observed at different epochs. Yet this variability is known to decrease in amplitude with wavelength, and as it is discussed in \citet{rob2008} the most extreme variables are reduced to amplitudes of a few tenths of a magnitude in the MIR. Even so, some dispersion due to this effect is expected in colour-colour and colour-magnitude diagrams.

\begin{table}
\caption{Details of the bandpasses used in the text.}
\label{wvl}
\centering
\begin{tabular}{c c c c}
\hline\hline
\noalign{\smallskip}
Label&Survey&$\lambda_\mathrm{iso}~(\mu m)$&$\Delta\lambda~(\mu m)$\\
\hline
\noalign{\smallskip}
$J$&2MASS&1.235&0.162\\
$H$&2MASS&1.662&0.251\\
$K_\mathrm{S}$&2MASS&2.159&0.262\\
$[W1]$&WISE&3.353&0.662\\
$[W2]$&WISE&4.603&1.042\\
$[W3]$&WISE&11.561&5.507\\
$[W4]$&WISE&22.088&4.101\\
\noalign{\smallskip}
\hline
\end{tabular}
\end{table}

\begin{figure}
\centering
\includegraphics[width=9cm]{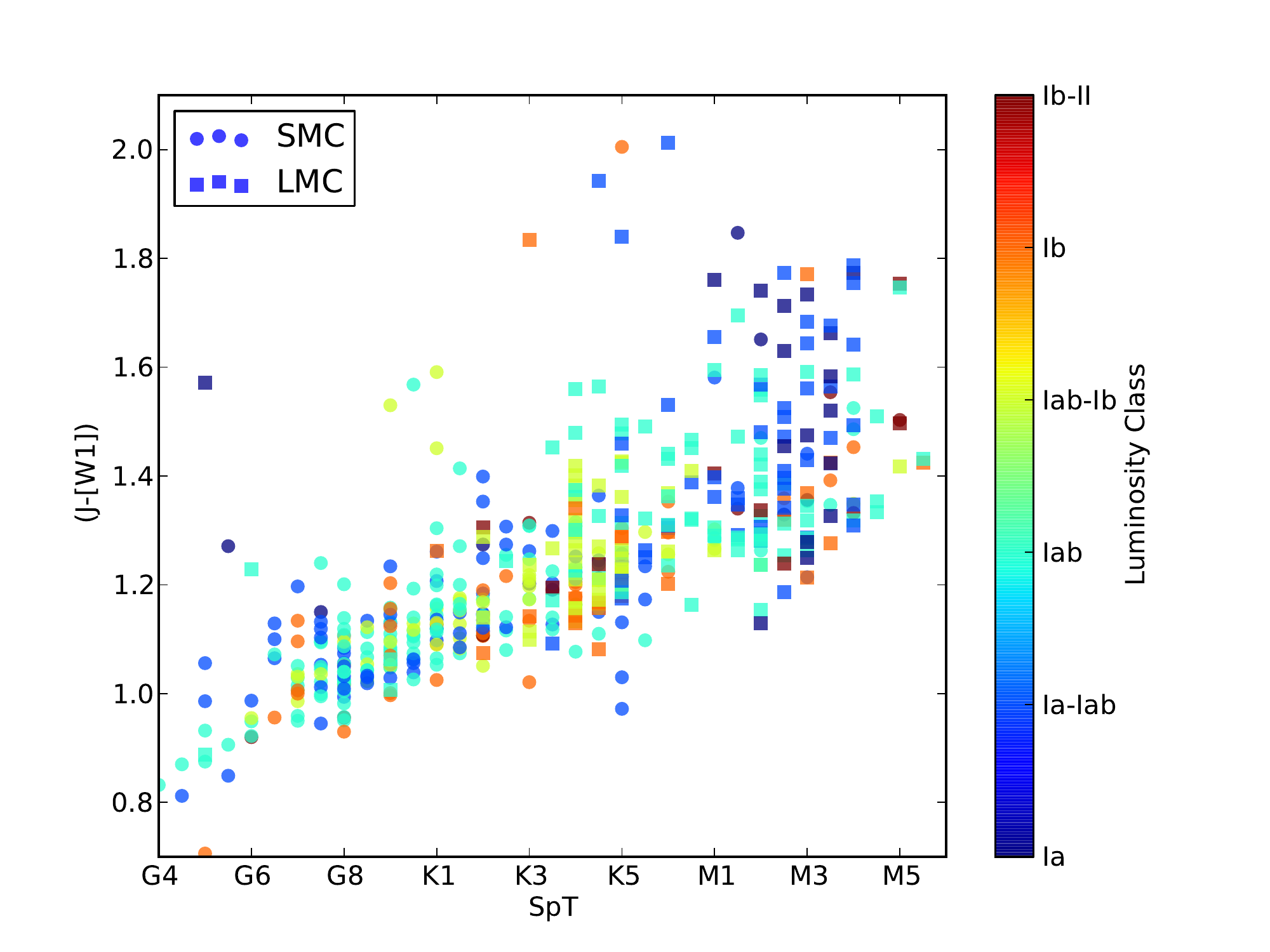}
\caption{Variation of $(J-[W1])$ with the spectral type of the RSGs in our sample.\label{jw1sp}}
\end{figure}

As has been mentioned, the magnitude of these RSGs in the $J$ band is affected by several atmospheric effects, such as molecular opacity and the appearance of dust, that are controlled by the effective temperature and surface gravity of the star. Interestingly, $[W1]$ is expected to be mostly photospheric and not subject to a strong absorption by the outer layers of the stellar envelope. The interplay of these factors results in the fact that the $(J-[W1])$ is a very good indicator of spectral type, as can be seen in Fig. \ref{jw1sp}. There is an almost linear relation between SpT and this colour. There are clear hints that this behaviour saturates around M3, but our sample lacks stars of later type, so we cannot confirm this fact.

\subsection{Dust, and mass loss}

\begin{figure*}
\centering
\includegraphics[width=9cm]{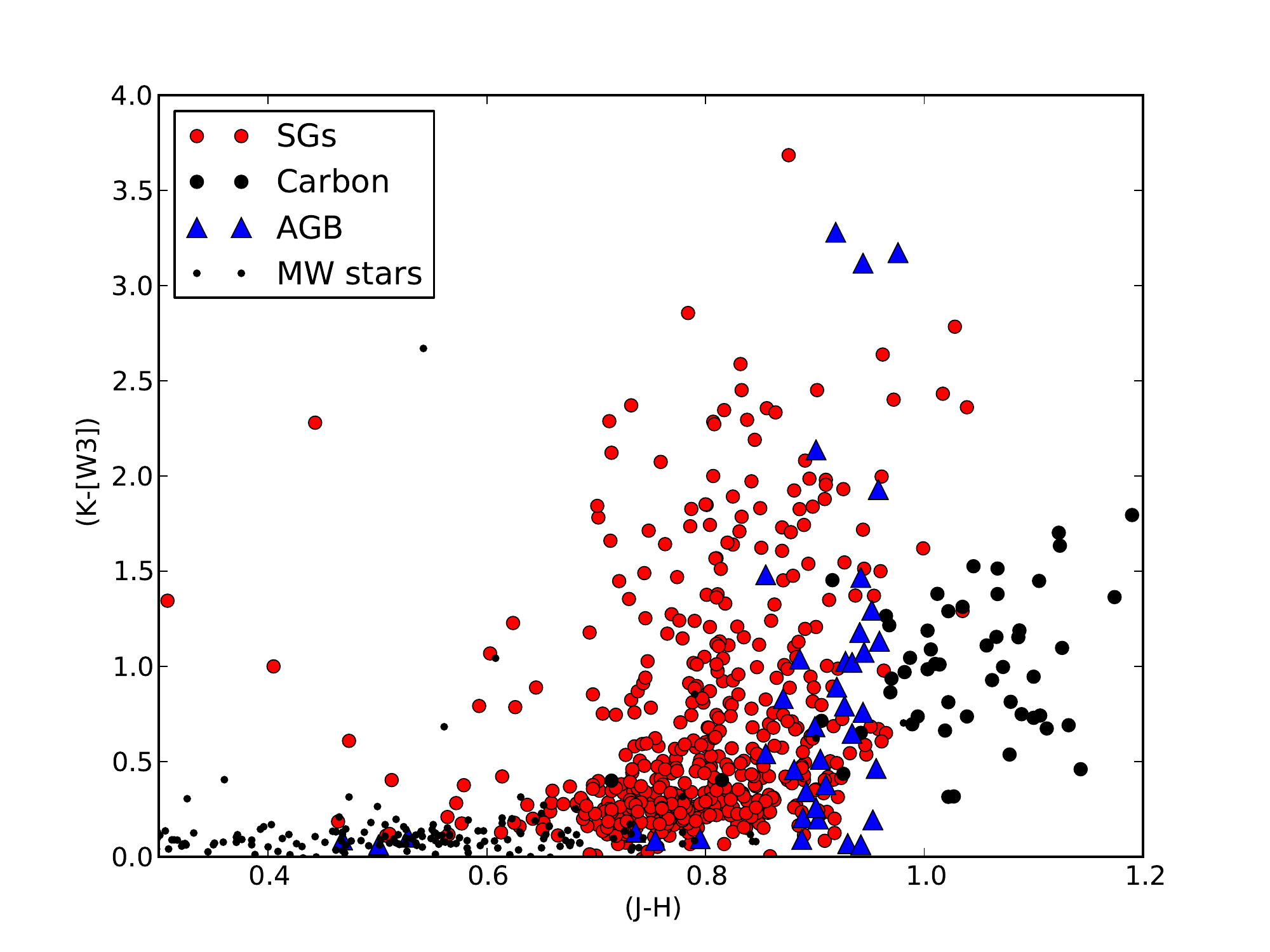}
\includegraphics[width=9cm]{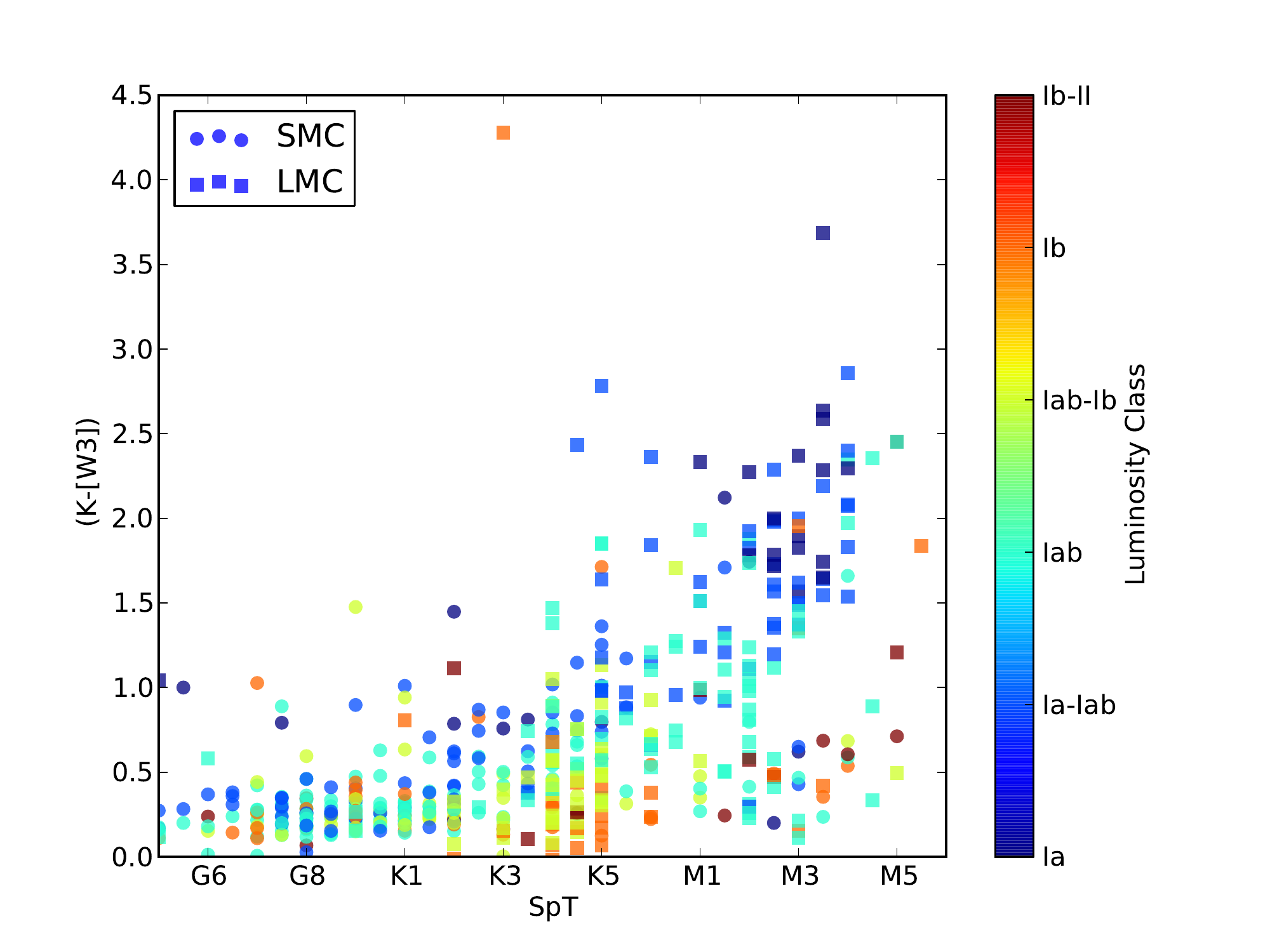}
\caption{{\bf Left:}$(J-H)$ vs. $(K_\mathrm{S}-[W3])$ diagram for the whole sample. {\bf Right:} $(K_\mathrm{S}-[W3]$) as a function of spectral type for all the SGs in the sample.\label{jhkw3}}
\end{figure*}

One of the most relevant physical phenomena affecting the atmospheres of late-type stars is that of mass loss. \citet{joss2000} show over IRAS data that the $(K_\mathrm{S}-[12])$ colour is a good measure of as mass loss, as this process will reflect on the MIR excess. The $W3$ band from WISE is the one that mimics most closely the IRAS $12\:\mu$m band, and so we explore the presence of mass loss using the $(K_\mathrm{S}-[W3])$ colour. As can be seen in Fig. \ref{jhkw3}, all of the MW population present in our sample falls in a stripe of $0\leq(K_\mathrm{S}-[W3])\leq0.3$. Almost all SGs, AGBs and carbon stars  present redder values of this colour.

Although the bulk of carbon stars tend to be redder than SGs (the majority of which satisfies $(K_\mathrm{S}-[W3])\leq 1.0$), these stars never reach values beyond $(K_\mathrm{S}-[W3]) = 2.0$, while several SGs and AGBs do so. This probably hints to an enhanced mass loss and the presence of strong winds. Although the mass loss in SGs is supposed to be much stronger than in AGBs, the distribution of the latter in this plot seems to follow closely that of late-type SGs. As due to our selection criteria we are only picking the bright end of the AGB population, this similarity points again -- even if with low statistical significance -- to the fact that there is not a sharp-cut transition from AGBs to RSGs. Similar effects have been observed by other authors, e.g. \citet{yang2012}.

The number of SGs showing evidences of mass loss becomes important around K5, but even for the most evolved types there are still stars that show small MIR excesses (and so potentially low mass loss). At all types, bright luminosity classes tend to feature higher values for $(K_\mathrm{S}-[W3])$.

\begin{figure}
\centering
\includegraphics[width=9cm]{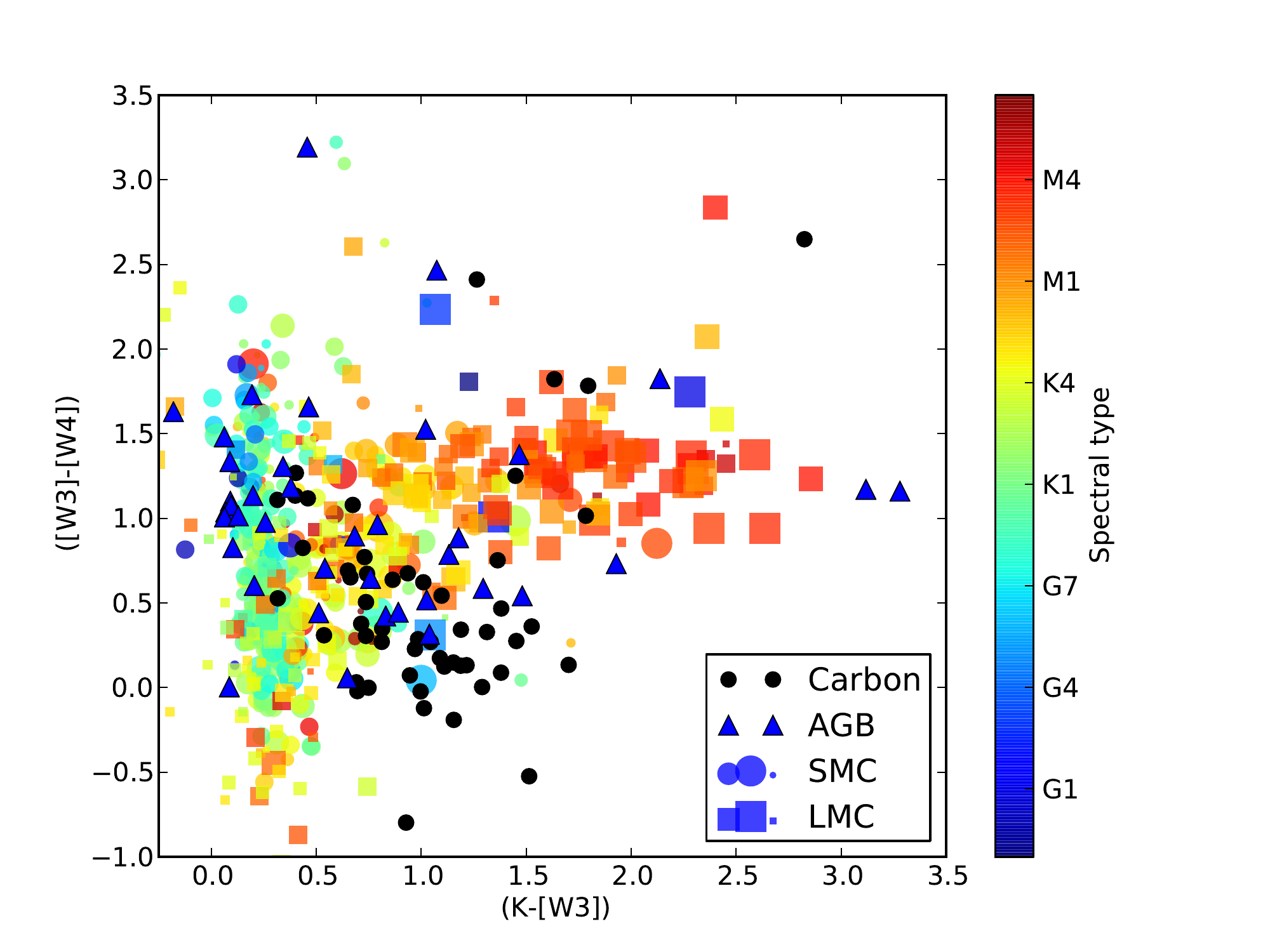}
\caption{Colour-colour diagram for our sample of MC stars, representing $(K_\mathrm{S}-[W3])$ (related to mass loss) against $([W3]-[W4])$, related to the temperature of the outer dust layers. In this plot, symbol size is a function of LC (i.e larger symbols imply brighter classes).}\label{kw3w3w4}
\end{figure}

The onset of dust in the outer layers of their atmospheres is one of the most relevant factors that dictate the photometric properties of cool, late type stars. Large granular compounds (complex carbon molecules, silicate grains, ice particles, etc.) form and coalesce in the outer layers of their extended atmospheres, and mediate their appearance in the NIR and MIR. The thermal emission from these particles becomes dominant at long wavelengths \citep{smith2001}. In Fig. \ref{kw3w3w4} we compare $([W3]-[W4])$, a colour that should be a good indicator of the dust temperature in the outer layers of the stellar shroud, with $(K_\mathrm{S}-[W3]$), that measures mass loss. As noted before only late-type SGs show mass loss, with a marked  tendency to higher values than AGBs or carbon stars.

As we can assign membership to the clouds to our stars, it is possible to estimate their intrinsic luminosities by assuming a given distance modulus to each cloud. We use $\mu=18.48$ for the LMC \citep{walker2012} and $\mu=18.99$ for the SMC \citep{grac2014}. Although not all the stars will be at the same distance, the spread of both clouds in distance modulus is low, and the dispersion introduced in the intrinsic magnitudes negligible for the kind of qualitative analysis we are going to
conduct here.

\begin{figure*}
\centering
\includegraphics[width=9cm]{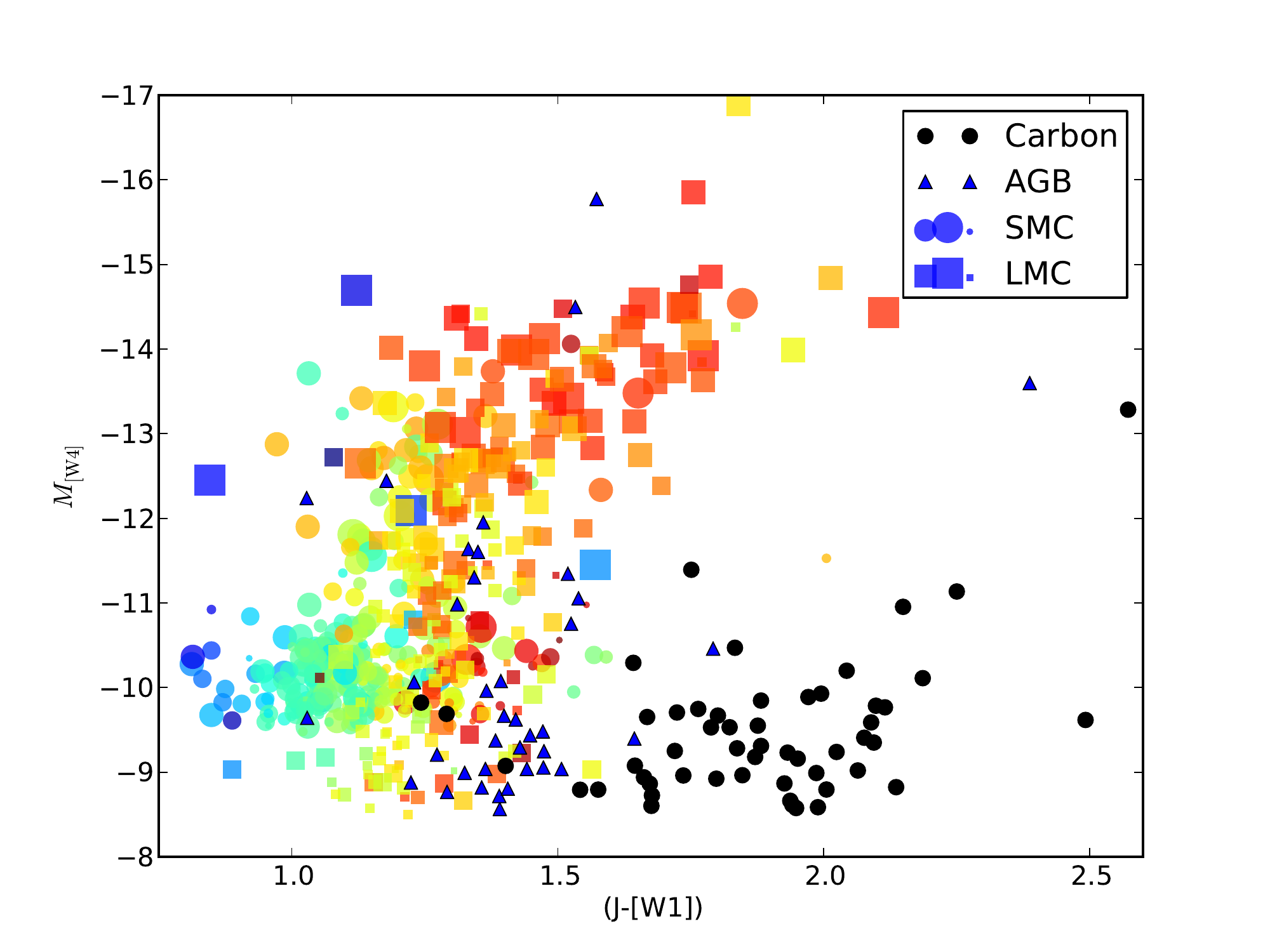}
\includegraphics[width=9cm]{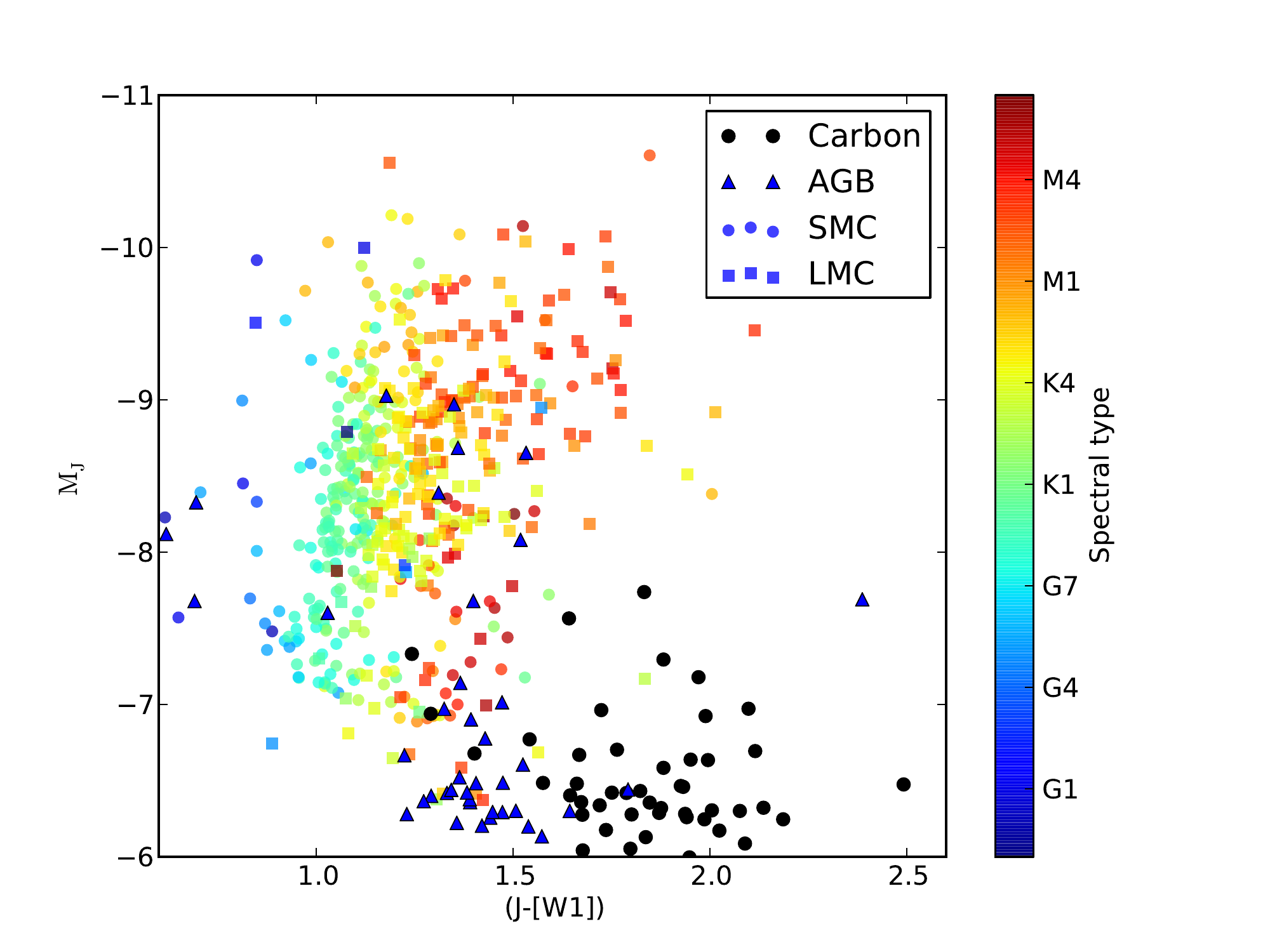}
\caption{Two different CMDs for our sample, both using the $(J-[W1])$ colour, that separates well the different populations. {\bf Left: } Absolute magnitude in $[W4]$, dominated by the outermost layers of the stellar envelope. In this plot, symbol size is a function of LC (i.e larger symbols imply brighter classes). {\bf Right:} Absolute magnitude in the $J$ band, for which the bolometric correction is lower.}\label{jw1w2w4}
\end{figure*}

In Fig. \ref{jw1w2w4} we plot two CMDs for our sample.

As expected (Fig. \ref{jw1w2w4}, right panel), there is a brightness limit for SGs, and no star in our sample goes beyond $M_\mathrm{J}\sim-10.5$, a limit that translates into $M_\mathrm{K_S}\sim-11.5$. Although the brightest stars in these NIR bands tend to be of late type, there is not a strong correlation between brightness and spectral type. As expected, AGBs and carbon stars occupy the lower end of the magnitude ladder, although some of the latter can reach $M_\mathrm{J}\sim-9$.

This is not the case for the brightness in the MIR. SGs with types earlier than K all cluster around $M_\mathrm{[W4]}\sim-10$. The K sequence of types is more or less evenly distributed between $-10\geq M_\mathrm{[W4]}\geq -13$. Finally, late type, M RSGs reach up to $M_\mathrm{[W4]}\sim-15$, with what appears to be a linear relation between absolute magnitude and colour, likely to arise from the increase of dust in their outer layer, as this would make the star dimmer in $J$ (hence redder) while brighter in $[W4]$, as this band is dominated by the thermal emission of dust. There are some stars of various types that break this upper limit in $M_\mathrm{[W4]}$, but the photometry of all of them seems to be affected by the nebular emission of large, nearby \ion{H}{ii} regions, such as 30 Doradus.

It is worth noting that both CMDs sample very different physical elements of the stars, as for objects with extended envelopes and strong losses, in the NIR infrared we are looking at the central object while in the red part of the MIR these extended atmospheres are the dominant component.

\subsection{Objects of particular interest}

\begin{itemize}

\item HV~838: This SMC star is a known large-amplitude photometric variable with a period of $\sim$660 days \citep{woo1983}. It has been considered both an RSG and an AGB  star in different works. We have observed it in all three SMC campaigns but only the blue spectrum from 2011 has a SNR sufficient to allow a proper classification. The spectrum shows strong H~line emission. In the infrared region, it displays strong inverse P-Cygni profiles in the Ca triplet and other nearby lines. This behaviour is completely atypical in RSGs.

In view of this, we decided to study the infrared spectra from 2010 and 2012. There are no emission lines in these, but the spectral types are very late, explaining why the blue spectra have such a low SNR. Based only in the infrared spectra, we  find types $\sim$M7\,II in 2010 and $\sim$M8\,II in 2012, but in 2011 the infrarred spectrum reveals a late K\,Ib star.

While this extreme change in spectral type is very unfrequent among RSGs, \cite{sos2011} found a variation of almost 3~magnitudes in the $I$-band, much larger than the typical variations for RSGs, $\triangle I\lesssim~0.5$~mag  \citep{gro2009}, and in concordance with the severe spectral variations.

In consequence, despite its high luminosity, $M_{K_{\mathrm S}}=-9.55$~mag, more typical of RSGs than AGB stars, we have to conclude that HV~838 is a very luminous AGB star. This is in agreement with its position in the period/luminosity diagram \citep{woo1983} and the presence of a significant \ion{Li}{i}~6707\AA\ line \citep{smi1995}.

\item HV~1956 ([M2002]~58738): This star is a well known long-period Cepheid variable \citep{but1976,egg1977}, with a very long period of $\sim210$ days \citep{sos2010}. This star has been classified as M0--1\,I by \citep{pre1983}, G2\,Ib by \cite{wal1984} and K2\,I by \cite{mas2002}. We have observed it in all three epochs, finding spectral types G5\,Ia in 2010, G7\,Ia in 2011 and G4\,Ia in 2012. This is a very peculiar object and while we cannot discard the possibility of very large spectral variations, we can conclude that it spends most of its time in the G spectral type, as expected for a Cepheid variable.

In addition, our red spectra show inverse P-Cygni emission for most of the lines in 2010, no emission in 2011, and Cepheid-like profiles for the Ca triplet lines only in 2012.

Finally, we have to note this star has a $(J-H)\sim0.4$. Therefore, in Fig.~\ref{nircolsp}, it is the only G~star with the expected values for giants (the solid red line).

\item SMC091~(PMMR~10): This star shows a low radial velocity ($v_{\mathrm{hel}}=54.5\:{\rm km}\,{\rm s}^{-1}$), a value well under our velocity threshold to be considered a member of the SMC. However, our classification for it is K4\,Ia. Its high luminosity cannot be doubted and therefore it is a RSG, as first suggested by \cite{pre1983}. It has $K_\mathrm{S}=9.062$~mag, typical of SMC RSGs (see Fig.~\ref{vl_lc}). Therefore, despite its anomalous velocity, it is an RSG at the distance of the SMC. We speculate that is may be a runaway, ejected from a cluster or, more likely, from a binary in a supernova explosion.

\item SMC121: This star presents very atypical MIR photometry, with low photometric errors. It has a $M_{[W4]}=-15.7$ and ($[W3]-[W4]$)$=4.4$, indicative of an extreme dust envelope. Both values are close to being the highest among all our RSGs and AGB stars. The spectrum of this star presents no anomalies, and we have classified it as G8\,Ib. Inspection of the WISE W4 images for the area show that the star is projected on a patch of extended bright emission, associated to the \ion{H}{ii} region LHA 115-N  23. The anomalous MIR colours are undoubtedly due to contamination by the extended emission. The nearby RSG SMC123 is also likely contaminated by emission from this \ion{H}{ii} region and/or LHA~115-N22. For this reason, both stars have been removed from the plots.

\item SMC145: Its NIR spectrum is similar to a Carbon Star, but the Ca triplet is still visible. Its optical spectrum is not that of a Carbon star either, but it does not look similar to any of our standard or reference stars. We speculate that it may be related to S~stars.

\item SMC169: This very late star displays an optical spectrum around M8.5\,III. However, it has a $M_\mathrm{K_{S}}=-10.2$~mag, more typical of a RSG than a giant star. \citet{sos2009} give a main period of 1062 days, with an amplitude in the $I$ band of 2.7~mag, much larger than the typical values for RSGs,   \citep[$\sim0.5$~mag;][]{gro2009}. We have also examined the NIR spectrum, which shows many metallic lines despite its late spectral type. Therefore, we consider it as very luminous AGB star, and we have classified it as M8\,II.

\item SMC283~(CM~Tuc): This is the brightest star in our sample, with $K_\mathrm{S}=4.785$~mag. It also has the only negative radial velocity and the latest SpT (M6\,Iab) in our sample. Attending to its velocity and brightness, it cannot belong to the SMC. In fact, it was previously identified as a foreground star because of its velocity by \cite{pre1983}.

In mid or late-M luminous stars, the rise of TiO bands erodes the continuum, weakening or erasing the atomic lines. In consequence, the exact luminosity class of this object cannot be ascertained. Being a foreground star, we have no information about its intrinsic luminosity. Therefore, even if morphologically it shows the properties of a supergiant, it may be an AGB star.

The average intrinsic $K_{\mathrm S}$ magnitude for other stars with similar spectral type (M5\,I) in our SMC sample is $M_\mathrm{K_{S}}=-10.2$~mag. If we consider this as the intrinsic magnitude for our star, we obtain for it a distance modulus of 14.5 mag, i.e.~10~kpc, too far from the SMC to have any relation with it. However, as a galactic star, the reason for this location and velocity remains without explanation.

\item SMC311 (HV~12149): This very late star shows a spectrum about M8\,III. However, it has a $M_\mathrm{K_{S}}=-10.4$~mag, more typical of RSGs than giant stars. \citet{sos2009} give a main period of 769 days, with an amplitude of 2.3~mag, much larger than the typical variations for RSGs. We have also examined the NIR spectrum, which shows many metallic lines despite its late spectral type. Therefore, we consider it as very luminous AGB star, and we have classified it as M8.5\,II.

\item SMC401 (HV 2112): With a spectral type about M5.5\,II, this object has $M_\mathrm{K_\mathrm{S}}=-10.3$~mag, again very bright for a giant. However, OGLE has classified it as an unresolved multiple star, perhaps explaining its atypically high brightness in the $K_\mathrm{S}$ band. In any case, this has to be a very luminous AGB star.

\item LMC039: Its velocity is higher than the assumed threshold for the LMC. Its spectrum, however, corresponds without a doubt to a RSG, while its brightness is inside the typical range for RSGs in the LMC. In consequence, we consider it as an LMC RSG with peculiar velocity.

\item LMC074~(HV~2572): This object was proposed by \cite{woo1985} as a candidate low-luminosity RSG, because of a photometric period of only 201 days. The spectroscopic data disprove this possibility. Our spectral type, combining the blue and infrared spectra, is M7\,III, while \citet{smi1995} report the detection of a strong \ion{Li}{i}~6707\AA\ line. Even though \cite{gro2009} report a photometric period of 312 days, OGLE~III data \citep{sos2009} give a main period of 605 days, more in line with the luminosity/period relation for AGB stars. Also,  \citet{sos2009} observed an amplitude in $I$ of $2.5$~mag, too large for a RSG \citep{gro2009}. HV~2572 is therefore a luminous AGB star, with $M_\mathrm{K_{S}}=-10.1$~mag (above the upper $M_\mathrm{K_{S}}$ limit in the period/luminosity diagram of \citealt{woo1983}). Therefore we have opted to classify it as M7.5\,II-III.

\item LMC169~(HV~2670): This star has a radial velocity below the threshold adopted, but its spectrum shows clear RSG features. Its brightness is also typical for a LMC RSG. Therefore, we consider that this star is a LMC RSG with peculiar velocity.

\end{itemize}

We have also found emission in Balmer H~lines for some of our RSGs. As they present no other typical nebular emission lines, we consider this emission as intrinsic. On the other side, in the IR spectra there are no emission lines or any other peculiarities. In the SMC, stars with Balmer line emission are: SMC374, SMC372, [S84d] 105-7, and [M2002]~SMC 8324, 8930, 9766, 13472, 18592, 23463 \& 55355. In the LMC sample, we find: LMC122 and [M2002]~LMC 143035 \& 148381.

A number of targets show evidence for a blue companion in their optical spectra. We list:

\begin{itemize}
\item $\left[ \mathrm{M}2002 \right]$~55933: The blue end of its optical spectrum is dominated by the signal of an early~B star. We have checked available images in the $U$ band and found a large number of blue objects around our star. Therefore, this early-B star might be a visual companion.

\item $\left[ \mathrm{M}2002 \right]$~67554: For wavelenghts shorter than 4300\AA\ the spectrum is dominated by the flux of an early-B star. In the available image in the $U$ band, there is a blue star $4\arcsec$ away. The possibility that some of the flux is collected by the fibre cannot be discarded, and so the blue star might be the visual companion.

\item $\left[ \mathrm{M}2002\right]$~51906: This star has an H$\delta$ absorption line stronger than usual for its SpT. It may be caused by a physical early-type companion, as no blue stars close to this RSG are seen in the $U$-band image.

\item YSG010: The blue end of its optical spectrum is dominated by the flux of a B star. Given the shape of its lines, it has to be a fast rotator. As in the $U$-band image there are not other blue stars close to it, the B star may be a binary companion. This object presents emission in the Balmer lines, but there are no nebular emission lines, nor emission features in the nIR spectrum.

\item $\left[ \mathrm{M}2002 \right]$~169142, 168047 and LMC238: In all cases, the blue end of the optical spectrum is dominated by the flux of an early~B star. As these stars are in the middle of clusters (H88~298, KMK88~91 and BSDL~2654, respectively) the B star is probably another spatially-separated member.

\item LMC172: The blue end of its optic spectrum shows the inprint of an early star. The $U$-band image does not provide definite information, and this star may be a visual companion. As the S/N is low for this possible companion, a more detailed classification is not possible.

\item LMC049: This star is in the middle of the cluster NGC~1967, and the blue end of its optical spectrum is dominated by the flux of a B0--B1 star.

\item LMC239: This star is in the middle of the cluster H88~301, and its blue spectrum shows traces of an early B star.

\item LMC256: This star is in the middle of the cluster H88~308, and its blue optic spectrum is dominated by a $\sim$B1\,III star.

\item SMC099: An early-B star appears in its blue spectrum. However, in the $U$~band image there are no bright sources close to this star. Therefore, this early-B contaminant may be a physical companion.

\item LMC062: This star is in the cluster NGC~1983. Its optical spectrum is completely dominated by an B9\,I star, but its red spectrum is a blend of a bright K-type supergiant and an earlier component, probably a bright G o F-type star.

\item LMC110: Despite the late spectral type of this star (M5\,Ib-II), it presents strong Balmer lines in the blue sectrum. Since no \ion{He}{i} lines are seen, this early-type companion may be a late-B or early-A star. In this case, its LC should be at least II to be observable in the LMC. There are no indications of a cluster close to this star.

\end{itemize}

\section{Conclusions}

We have performed a pilot study in Large and Small Magellanic Clouds, aimed at their red supergiant population. Over a set of photometrically selected candidates, we have performed a detailed spectroscopic analysis, deriving spectral types, luminosity classes and line-of-sight velocities for all the observed targets. Once classified and with the available photometry, we show that:
\begin{itemize}
\item[-] There is a large population of supergiants in both clouds, largely in the dim end of their brightness range, that remains to be observed.
\item[-] There is no purely photometric criterion capable of separating completely different populations, and when  selecting RSGs, we will always have to choose between the completeness of the photometric sample and its cleanliness. Due to the fact that instead of clear-cut borders, the transitions between different populations are gradual, there will always be some AGB and carbon stars that will appear as interlopers.
\item[-] It is possible, nonetheless, to use the synergies between near and mid infrared photometry  to open avenues to much more efficient selection criteria.
\item[-] The completeness of the photometric selection criteria is a function of spectral type, and in particular there is a loss of efficiency for red supergiants with the earliest and latest types. This has to be weighed in whenever drawing conclusions about their relatives abundances in several astrophysical contexts.
\item[-] Mass loss becomes important only for supergiants later than K5, although it is not ubiquitous and at each spectral type there will stars with no or very little apparent mass loss.
\item[-] The thermal behaviour of the dust inhabiting these expelled outer layers seems to be similar for all the supergiants, independent of spectral type and luminosity class.
\end{itemize}

\begin{acknowledgements}
The observations have been supported by the OPTICON project (observing proposals 2010B/01, 2011A/014 and 2012A/015), which is funded by the European Commission under the Seventh Framework Programme (FP7).
Part of the observations have been taken under service mode, (service proposal AO171) and the authors gratefully acknowledge the help of the AAO support astronomers.
This research is partially supported by the Spanish Ministerio de Econom\'{\i}a y Competitividad (Mineco) under grant AYA2012-39364-C02-02.
The work reported on in this publication has been partially supported by the European Science Foundation (ESF), in the framework of the GREAT Research Networking Programme.
This research was achieved using the POLLUX database (http://pollux.graal.univ-montp2.fr) operated at LUPM  (Université Montpellier II - CNRS, France with the support of the PNPS and INSU. This research made use of the Simbad, Vizier, and Aladin services developed at the Centre de Donn\'ees Astronomiques de Strasbourg, France.
This publication makes use of data products from the Two Micron All Sky Survey, which is a joint project of the University of Massachusetts and the Infrared Processing and Analysis Center/California Institute of Technology, funded by the National Aeronautics and Space Administration and the National Science Foundation.
This publication makes use of data products from the Wide-field Infrared Survey Explorer, which is a joint project of the University of California, Los Angeles, and the Jet Propulsion Laboratory/California Institute of Technology, funded by the National Aeronautics and Space Administration.
\end{acknowledgements}

\bibliographystyle{aa}
\bibliography{general}

\appendix

\onecolumn

\section{Catalogue of observed sources}

{\setlength{\tabcolsep}{0.4em}
\begin{longtab}
\begin{landscape}
\begin{longtable}{c c | c c | c c c c c | c c c c | c c c | c c c c}
\caption{Summary of the observations\label{totalcat}}\\
\hline\hline
\noalign{\smallskip}
&&&&&&&&&\multicolumn{4}{c}{Epochs}&\multicolumn{3}{c}{2MASS}&\multicolumn{4}{c}{WISE}\\
ID$^1$&Cloud&RA&Dec&Origin$^2$&Var.&$v_\mathrm{HEL}$&SpT&LC$^3$&2010&2011&2012&2013&J&H&$\mathrm{K_S}$&W1&W2&W3&W4\\
\noalign{\smallskip}
\hline
\noalign{\smallskip}
\endfirsthead
\caption{continued.}\\
\noalign{\smallskip}
&&&&&&&&&\multicolumn{4}{c}{Epochs}&\multicolumn{3}{c}{2MASS}&\multicolumn{4}{c}{WISE}\\
ID$^1$&Cloud&RA&Dec&Origin$^2$&Var.$^3$&$v_\mathrm{HEL}$&SpT&LC$^4$&2010&2011&2012&2013&J&H&$\mathrm{K_S}$&W1&W2&W3&W4\\
\noalign{\smallskip}
\hline
\noalign{\smallskip}
\endhead
\noalign{\smallskip}
\hline
\endfoot
\noalign{\smallskip}
\multicolumn{20}{l}{(1) For stars taht show spectral variability, different epochs are recorded as different entries.}\\
\noalign{\smallskip}
\multicolumn{20}{l}{(2) "GDN" labels our photometrically selected sample, "Massey" are taken from \citet{mas2003} and references therein, and "YSG" from \citet{neu2010}.}\\
\noalign{\smallskip}
\multicolumn{20}{l}{(3) This column marks stars that show variability in our spectra, and so for stars with a single epoch is left blank.}\\
\noalign{\smallskip}
\multicolumn{20}{l}{(4) Luminosity class "C" is reserved for Carbon stars.}\\
\endlastfoot
LMC001    &LMC&80.21066667&-69.40222222&GDN  &-- &245.5&Unk  &C  &0&0&0&1&12.48&11.46&10.99&10.54&10.67&10.17&9.90\\
LMC002    &LMC&80.26645833&-69.45668056&GDN  &-- &268.8&M3   &Ib &0&0&0&1&11.43&10.60&10.34&10.22&10.42&10.19&9.77\\
LMC003    &LMC&80.34245833&-69.51676111&GDN  &-- &269.6&Unk  &C  &0&0&0&1&11.56&10.49&10.02&9.57&9.62&9.10&9.89\\
LMC004    &LMC&80.38912500&-69.20932778&GDN  &-- &271.9&Unk  &C  &0&0&0&1&12.31&11.23&10.70&--&--&--&--\\
LMC005    &LMC&80.47033333&-69.75530278&GDN  &-- &240.0&Unk  &C  &0&0&0&1&12.12&11.02&10.53&10.28&10.35&9.59&9.52\\
LMC006    &LMC&80.54104167&-69.79196111&GDN  &-- &283.7&M2.5 &II-III&0&0&0&1&12.12&11.22&10.87&10.73&10.79&10.36&9.92\\
LMC007    &LMC&80.59904167&-69.38114167&GDN  &-- &268.2&K4.5 &Iab-Ib&0&0&0&1&10.34&9.55&9.30&9.17&9.38&9.15&9.32\\
LMC008    &LMC&80.68191667&-69.62240556&GDN  &-- &287.2&K5   &Ib &0&0&0&1&10.48&9.67&9.39&9.27&9.49&9.32&9.98\\
\hline
\end{longtable}
\end{landscape}
\end{longtab}
}

\end{document}